\documentclass{aastex}
\usepackage{emulateapj5}

\def\cm#1{\ifmmode {\,{\rm cm^{-#1}}}                  
        \else \hbox{$\,${\rm cm$^{\rm -#1}$}}\fi}
\def\raw {\ifmmode\rightarrow\else$\rightarrow$\fi}
\def\ex#1{\ifmmode {\times 10^{#1}}         
        \else \hbox{{$\times 10^{\rm #1}$}}\fi}

\newcommand{\kms}{\mbox{km~s$^{-1}$}}
\newcommand{\mloss}{\mbox{$\dot{M}$}}
\newcommand{\my}{\mbox{$M_{\odot}$~yr$^{-1}$}}
\newcommand{\gsim}{\raisebox{-.4ex}{$\stackrel{>}{\scriptstyle \sim}$}}
\newcommand{\lsim}{\raisebox{-.4ex}{$\stackrel{<}{\scriptstyle \sim}$}}
\newcommand{\ls}{\mbox{$L_{\odot}$}}
\newcommand{\ms}{\mbox{$M_{\odot}$}}
\newcommand{\bri}{\mbox{erg\,s$^{-1}$\,cm$^{-2}$\,\AA$^{-1}$\,arcsec$^{-2}$}}
\newcommand{\brib}{\mbox{erg\,s$^{-1}$\,cm$^{-2}$\,arcsec$^{-2}$}}
\newcommand{\flux}{\mbox{erg\,s$^{-1}$\,cm$^{-2}$\,\AA$^{-1}$}}

\newcommand{\n}{\~n}

\newcommand{\crl}{CRL\,618}
\newcommand{\cs}{CS82}
\newcommand{\hrh}{HRH87}
\newcommand{\ai}{\mbox{\'{\i}}} 
\newcommand{\vsys}{\mbox{$V_{\rm sys}$}} 
\newcommand{\h}{$^{\rm h}$}
\newcommand{\m}{$^{\rm m}$}

\newcommand{\mi}{\mbox{$'$}}

\shorttitle{The proto-planetary nebula CRL 618. I. Optical spectroscopy and imaging}
\shortauthors{S\'anchez Contreras et al.}

\begin{document}


\title{Physical structure of the protoplanetary nebula
CRL\,618. I. Optical long-slit spectroscopy and imaging}


\author{C. S\'anchez Contreras, R. Sahai}
\affil{Jet Propulsion Laboratory, California Institute of Technology, 
MS 183-900, 4800 Oak Grove Drive, Pasadena, CA 91109}
\email{sanchez,sahai@eclipse.jpl.nasa.gov}

\author{A. Gil de Paz}
\affil{NASA/IPAC Extragalactic Database, California Institute of Technology,
MS 100-22, Pasadena, CA 91125}
\email{agpaz@ipac.caltech.edu}

\begin{abstract}
In this paper (paper I) we present 
optical long-slit spectroscopy and imaging of the
protoplanetary nebula \objectname{CRL 618}. The optical lobes of \crl\
consist of shock-excited gas, which emits many recombination and
forbidden lines, and dust, which scatters light from the innermost
regions. From the analysis of the scattered H$\alpha$ emission, we
derive a nebular inclination of $i$=24\degr$\pm$6\degr. The
spectrum of the innermost part of the east lobe (visible as a bright,
compact nebulosity close to the star in the H$\alpha$ $HST$ image) is
remarkably different from that of the shocked lobes but similar to
that of the inner \ion{H}{2} region, suggesting that this region
represents the outermost parts of the latter. We find a non-linear
radial variation of the gas velocity along the lobes.
The largest projected LSR velocities ($\sim$\,80\,\kms) are measured
at the tips of the lobes, where the direct images show the presence of
compact bow-shaped structures. The velocity of the shocks in
\crl\ is in the range $\sim$75-200\,\kms, as derived from diagnostic line
ratios and line profiles. We report a brightening (weakening) of
[\ion{O}{3}]$\lambda$5007\AA\ ([\ion{O}{1}]$\lambda$6300\AA) over the
last $\sim$\,10 years that may indicate a recent increase in the speed
of the exciting shocks. From the analysis of the spatial variation of
the nebular extinction, we find a large density contrast between the
material inside the lobes and beyond them: the optical lobes seem to
be `cavities' excavated in the AGB envelope by interaction with a more
tenuous post-AGB wind. The electron density, with a mean value $n_{\rm
e}$\,$\sim$5$\times$10$^3$-10$^4$\,\cm3, shows significant
fluctuations but no systematic decrease along the lobes, in agreement
with most line emission arising in a thin shell of shocked material
(the lobe walls) rather than in the post-AGB wind filling the interior of the
lobes. The masses of atomic and ionized gas, respectively, in the east
(west) lobe are $>$1.3$\times$10$^{-4}$\,\ms\
($>$7$\times$10$^{-5}$\,\ms) and $\sim$6$\times$10$^{-5}$\,\ms\
($\sim$4$\times$10$^{-5}$\,\ms). The shocks in \crl\ are in a
radiative regime and may lead in the future to the evolution of the
optically-emitting lobes into a fast, bipolar molecular outflow. The
time required by the dense, shocked gas to cool down significantly is
$\lsim$\,2\,yr, which is substantially lower than the kinematical age
of the lobes ($\lsim$\,180\,yr). This result suggests that a fast wind
is currently active in \crl\ and keeps shocking the circumstellar
material.
\end{abstract}



\section{Introduction}
\label{intro}
The physical mechanisms responsible for the onset of bipolarity and polar
acceleration in planetary nebulae (PNe) are already active
in the first stages of the evolution beyond the Asymptotic Giant
Branch (AGB), i.e$.$ in proto-Planetary Nebulae (PPNe, also called
post-AGB objects). Therefore, PPNe and young PNe hold the key for
understanding the complex and fast ($\sim$\,10$^3$\,yr) nebular
evolution from the AGB towards the PN phase. Such evolution is
believed (by an increasing number of astronomers) to be governed by
the interaction between fast, collimated winds or jets, ejected in
the late-AGB or early post-AGB phase, and the spherical and slowly
expanding circumstellar envelope (CSE) resulted from the star 
mass-loss process during the AGB \citep[see][]{sah98,kas00}.

\objectname{CRL 618} (= \objectname{RAFGL 618} = 
\objectname{IRAS 04395+3601} = \objectname{Westbrook Nebula}) 
is a well studied PPN which has very recently started its post-AGB journey
\citep[only $\sim$\,200\,yr ago, e.g.][]{kwo84} and is rapidly
evolving towards the PN stage.  Most of the circumstellar matter in
\crl\ is still in the form of molecular gas. This component, with a
total mass of $\approx$\,1.5\ms, consists of: (1) a spherical and
extended ($\gsim$\,20\arcsec) envelope expanding at 17.5\,\kms\
\citep{kna85,buj88,buj01,bac88,haj96,phi92}; and (2) an inner, compact
bipolar outflow moving away from the star at velocities
$\gsim$\,200\,\kms\ \citep{cer89,gam89,mei98,ner92}. The outer, and
slowly expanding component is interpreted as the result from the
mass-loss process of the central star during the AGB, which took place
at a rate of \mloss\,=\,few$\times$\,10$^{-5}$-10$^{-4}$\,\my. The
fast bipolar outflow, with a mass of $\sim$10$^{-2}$\ms, is believed
to be the result of the interaction between a fast, collimated
post-AGB wind and the spherical AGB CSE (see references above).

The high-excitation nebula (atomic and ionized gas) is composed of:
(1) a compact \ion{H}{2} region, visible through radio-continuum
emission, elongated in the E-W direction with an angular size
of 0\farcs2-0\farcs4 that is increasing with time
\citep{kwo81,mar93};  and (2) multiple lobes with shock-excited 
gas which produces recombination and forbidden line emission in the
optical \citep[e.g.][]{goo91,tra93,kel92,tra00}.
From previous spectroscopic data, the inner \ion{H}{2} region and the lobes 
are known to be expanding with
velocities of $\sim$\,20\,\kms\ and up to $\sim$\,120\,\kms,
respectively \citep{mar88,car82}. The analysis of different optical
line ratios indicates that a relatively large range of temperatures
($\approx$\,10,000 to 25,000\,K) and densities ($\approx$\,1600 to
8000\,\cm3) are present in the lobes \citep{kel92}.

The optical spectrum of \crl\ also shows a weak, red continuum which
is the stellar light reflected by the nebular dust.
From spectropolarimetric observations it is known
that a fraction of the flux of the Balmer lines is also scattered
light originally arising from the inner, compact \ion{H}{2} region
\citep{sch81,tra93}. The polarization of the forbidden lines is
negligible, indicating that they are almost entirely produced by the
shock-excited gas in the lobes with a small or insignificant
contribution by scattered photons from the \ion{H}{2} region (the high
density in this region, $\sim$10$^6$\,\cm3, produces collisional
de-excitation of most forbidden lines).

The central star of CRL\,618 has been classified as B0 based
on the shape of the dereddened visual continuum \citep{sch81} and on
the weak [\ion{O}{3}] line emission from the inner, compact \ion{H}{2}
region \citep{kal78}. The luminosity of CRL\,618, obtained by integrating
the observed IRAS fluxes, is $L=1.22 \times 10^4$\ls$[D/{\rm kpc}]^2$
\citep{goo91}. Based on the $L$ values predicted by
evolutionary models ($\sim$\,10$^4$\ls) and the typical scale
height of PN ($\sim$\,120\,pc) these authors calculate a distance to
the source of 0.9\,kpc, which we also assume in this paper.

In this paper (paper I of a series of two) we report optical imaging
and long-slit spectroscopy of
\crl. Observational techniques and data reduction are 
described in Sect$.$ \ref{obs}. 
In Sections \ref{morpho} and \ref{sect4} we present our 
observations, including a brief description of the
nebular morphology and main characteristics of the optical
spectrum. In Sect$.$
\ref{sect5} we analyze the different emission line components
and study the physical properties of the different nebular regions
probed by them. In Sects$.$ \ref{incli} and \ref{str+kin} we derive
the mean nebular inclination and describe the kinematics of \crl,
respectively. The spatial distribution along the nebular axis of the
extinction and the electron density are presented, respectively, in
Sects$.$ \ref{sext} and \ref{physical}. In the latter, we also
estimate the atomic and ionized mass in different regions. Finally, 
we discuss our results and give a possible scenario to explain the
formation and future evolution of \crl\ in Sect$.$ \ref{discuss}. The
main conclusions of this work are summarized in Sect$.$
\ref{conc}.

\section{Observations and data reduction}
\label{obs}

\subsection{Optical imaging}
\label{imaging}

\subsubsection{Ground-based}
We have obtained narrow-band images of the H$\alpha$ line and the
adjacent continuum emission in the PPN \crl\
(Fig$.$\,\ref{f1}a,b). Observations were carried out on November
8$^{\rm th}$ and 9$^{\rm th}$, 2001 with the Palomar Observatory
60-inch telescope. The detector was the CCD camera \#13, which has
2048$\times$2048 pixels of 24$\mu$m in size and provides a plate scale
of 0\farcs378\,pixel$^{-1}$. The narrow-band filters used for imaging
the H$\alpha$ line and adjacent continuum are centered at 6563 and
6616\,\AA, respectively, and have a full width at half maximum (FWHM)
of 20\,\AA. Weather conditions during the observations were good,
although non-photometric, and the seeing ranged between 0\farcs9 and
2\arcsec.  The final spatial resolution of the H$\alpha$ (continuum)
combined images is 1\farcs9 (1\farcs1).  Total exposure times
are 2700\,s for the H$\alpha$ image and 6300\,s for the continuum.

Data reduction, including bias subtraction, flat field correction,
removal of cosmic rays and bad pixels, and registering+combination of
individual images, was performed following standard procedures within
IRAF\footnote{IRAF is distributed by the National Optical Astronomy
Observatories, which are operated by the Association of Universities
for Research in Astronomy, Inc., under cooperative agreement with the
National Science Foundation.}.

Astrometry of our images was done cross-correlating
the coordinates of 376 and 187 field stars (for H$\alpha$ and the
continuum, respectively) in the CCD and those in the USNO-A2.0
catalog. This procedure yields a plate scale of
0\farcs378$\pm$0\farcs009\,pixel$^{-1}$. 
The coordinates of the origin of our spatial scale in our images 
are R.A.= 04\h42\m53\fs55 and Dec.= 36\degr06\mi54\farcs5 (J2000). 
The errors of the absolute
positions derived are $\lsim$\,0\farcs4 (considering the standard
deviation of our best calibration solution, 0\farcs15, and the
absolute position error of the USNO stars, about 0\farcs25).

We have obtained a pure H$\alpha$ line emission image by subtracting
the continuum emission to our original H$\alpha$+continuum image. The
continuum image was first smoothed to match the lower effective
spatial resolution of the H$\alpha$ image and then the flux of several
field stars was measured in both images to derive a scaling factor
that includes the differences of the system `filter+CCD' response
between the two images as well as the different exposure times and sky
transparency when the observations were performed.

Flux calibration was performed using the USNO-A2.0 $R$ magnitudes of
the field stars.  The error in the flux calibration is dominated by
the absolute flux error in the USNO-A2.0 catalog ($\sim$55\%).  The
flux calibration of the H$\alpha$ and continuum images is consistent
with our calibrated long-slit spectra within a factor $\lsim$\,2.

\subsubsection{Hubble Space Telescope}

As a complement to our data, we have used the high angular resolution
H$\alpha$ image of \crl\ from the Hubble Space Telescope ($HST$)
archive (GO 6761, PI: S. Trammell). This image was obtained with the
Wide Field Planetary Camera 2 (WFPC2) and the narrow-band filter
F656N, around the H$\alpha$ line.  The WFPC2 has a
36\arcsec$\times$36\arcsec\ field of view and a plate scale of
0\farcs0455 pixel$^{-1}$. The image (Fig$.$\,\ref{f1}c) is
pipeline-reduced.



\subsection{Optical long-slit spectroscopy}
\label{obsop}
We obtained optical long-slit spectra of CRL\,618 on November 12$^{\rm
th}$ and 13$^{\rm th}$ 2000, using the Intermediate Dispersion
Spectrograph (IDS) of the 2.5m Isaac Newton Telescope of the Roque de
los Muchachos Observatory (La Palma, Spain). The detector was the
EEV10 CCD, with squared pixels of 13.5$\mu$m lateral size. Only a
clear and unvignetted region of 700$\times$2600 pixels was used (the
2600 pixels were along the spectral axis). The CCD was mounted on the
500mm camera, leading to a spatial scale of
0\farcs19\,pixel$^{-1}$. The R1200Y and R900V gratings were employed
providing spectra in the (6233-6806)\AA\ and (4558-5331)\AA\
wavelength ranges with dispersions of 16.4\,\AA mm$^{-1}$ (0.22\,\AA
pixel$^{-1}$) and 22.1\,\AA mm$^{-1}$ (0.29\,\AA pixel$^{-1}$),
respectively.

A total of three slit positions were observed at position angles (PA)
of 93\degr\ (along the nebular symmetry axis) and 3\degr\ (through the
west lobe of the nebula). The slit positions are shown in Fig$.$\,\ref{f2}
superimposed on the H$\alpha$ ground-based image of CRL\,618.
At PA=93\degr\ two spectra were obtained: one with the slit placed on
the northernmost pair of lobes (hereafter, slit N93) and the other one
with the slit displaced $\sim$\,1\arcsec\ towards the South (partially
covering the southernmost lobes; slit S93).  At position N93, spectra
with two gratings, R1200Y and R900V, were obtained with exposures
times of 2000\,s and 2700\,s, respectively. For the
other two slit positions (S93 and PA=3\degr) only the R1200Y was used,
and exposure times were 2000\,s and 1300\,s respectively.
The slit was 1\arcsec\ wide and long enough to cover the whole nebula
(and a significant region of the sky).

The data were reduced following standard procedures for long-slit
spectroscopy within the IRAF package, including bias subtraction,
flat-fielding and illumination corrections, sky subtraction, and
cosmic-ray rejection. We used CuNe and CuAr lamps to perform the
wavelength calibration. The spectral resolution achieved (FWHM of the
calibration lamp lines) is $\sim$\,50\,\kms, at H$\alpha$, and
$\sim$\,80\,\kms, at H$\beta$.  Flux calibration was done using
sensitivity functions derived from two spectrophotometric standards,
HR1544 and HD217086, observed at different airmasses, and taking into
account the atmospheric extinction curve at La Palma. The geometric
distortion of the long-slit spectra ($\sim$\,0.06\%) was corrected
also using one of the flux calibration stars. The effective spatial
resolution obtained ranges between $\sim$\,1\arcsec\ and 1\farcs6.


\section{Morphology of the optical nebula}
\label{morpho}


In Fig$.$\,\ref{f1} we show direct images of the PPN \crl\ in the
light of the 6616\,\AA\ continuum (a) and H$\alpha$ line emission (b
and c) obtained from the ground and with $HST$. 
The H$\alpha$ images show the shocked, optical lobes of \crl\ consist
of several components with a very rich structure at a scale of
$\sim$\,0\farcs2 (see Fig$.$\,\ref{f1}c). The east and west lobes are
both composed of several, collimated jet-like structures emanating
from the center of the nebula. Bright, bow-shaped ripple-like features
are frequent along the lobes.


In the H$\alpha$ images of CRL\,618 we have labeled A' the small,
bright region of the east lobe closer to the nebular center
(Fig$.$\,\ref{f1}c). This region is very bright in the H$\alpha$ image
but very weak, e.g., in the WFPC2/$HST$
[\ion{O}{1}]$\lambda$6300\,\AA\ and
[\ion{S}{2}]$\lambda$$\lambda$6716,6731\,\AA\ images (see Trammell 2000; 
Sahai et al$.$ 2002, in preparation). As we will see in the
following sections, the spectrum of region A' is in fact substantially
different from that of the shock-excited lobes. This leads us to think
that region A' has a different nature, which we discuss in
\S\,\ref{spdist}.

A weak H$\alpha$ emission halo surrounds the lobes of \crl\
(Fig$.$\,\ref{f1}b,c). In the $HST$ H$\alpha$ image the halo is
observed very close to the boundaries of the emission lobes and is
particularly intense and extended in the east lobe: note the clumps of
diffuse emission ahead of this lobe (i.e$.$ just beyond the tip of the
lobe). The diffuse halo, like region A', is absent in the [\ion{O}{1}]
and [\ion{S}{2}] $HST$ images (see references above). In our
ground-based H$\alpha$ image (with higher signal-to-noise ratio than
that obtained with the $HST$) the size at a 3$\sigma$ level of the
halo is $\sim$\,35\arcsec$\times$22\arcsec, with the major axis
roughly oriented at PA=3\degr\ and centered at the position where the
maximum H$\alpha$ emission is observed. 

The continuum image of \crl\ traces the distribution of the dust
particles in the nebula (which reflect the stellar light) as well as
any field star. The central star of CRL\,618 is not directly visible,
which indicates that the stellar light is strongly obscured along the
line of sight. The continuum brightness distribution 
is in general similar to that of H$\alpha$, suggesting the 
presence of dust in the lobes. 

\section{The optical spectrum of CRL 618}
\label{sect4}


The optical spectrum of CRL\,618 is composed of recombination and
forbidden line emission superimposed on a faint, red continuum. 

\subsection{Long-slit spectra of emission lines} 
\label{splines}

In Fig$.$\,\ref{f2} we show long-slit spectra for the most intense
lines in the $\sim$\,[6230--6805]\,\AA\ range for slits N93, S93, and
PA3. The weakest lines detected in this range,
[\ion{S}{3}]$\lambda$6312.1\,\AA\ and \ion{He}{2}$\lambda$6678\,\AA,
for slit N93 are shown in Fig$.$\,\ref{f3}.
These spectra have been smoothed with a flat-topped rectangular kernel
of dimension 3$\times$3 pixels: the resulting degradation in the 
spatial and spectral resolution
is less than $\sim$\,4\% from the nominal value (see
\S\ref{obsop}) since the smoothing window was smaller than the local
seeing and FWHM of the calibration lines. The origin of the spatial
scale in the spectra (and in the images to the left and
in Fig$.$\,\ref{f1}) coincides with the point of maximum extinction
(measured as the highest H$\alpha$/H$\beta$ line ratio;
\S\,\ref{sext}). We will refer to this point also as to nebular center.
The LSR systemic velocity of the source, 
\vsys\,=--21.5$\pm$0.5\,\kms\ \citep[derived from molecular line emission
observations, e.g$.$,][]{cer89} is indicated by a vertical
line on each spectrum.

We find remarkable differences between the profile of the H$\alpha$
line (and also H$\beta$, see below) and those of most forbidden lines
for each slit position. These differences are specially noticeable in
the bright east lobe. A number of spectral
features\footnote{``Feature'' refers to a component in our
two-dimensional spectra.} (labeled A, B, and C in Fig$.$\,\ref{f4})
are seen to be superimposed on an underlying H$\alpha$ profile very
similar to that of [\ion{O}{1}]$\lambda$6300\,\AA\ (and most forbidden
lines). (Carsenty \& Solf, 1982, noted also the presence of 
two different components in the H$\alpha$ profile, see \S\,\ref{features}).

Feature A is the intense H$\alpha$ emission component observed in the
innermost region (closest to the central star) of the east lobe. The
maximum emission from this feature occurs at offset
$\sim$1\farcs4. For slit N93 such maximum is slightly red-shifted
($\sim$\,3.5\,\kms) with respect to \vsys\, whereas for S93, feature A
is blue-shifted by $\sim$\,10\,\kms. From comparison with the
H$\alpha$ images of \crl\ we determine that feature A originates in
the inner region A' (Fig$.$\,\ref{f1}). There is also some
blue-shifted [\ion{N}{2}]$\lambda\lambda$6548,6584\,\AA,
[\ion{S}{3}]$\lambda$6312\,\AA\ and \ion{He}{2}$\lambda$6678\,\AA\
emission arising in this region, which we refer to as emission feature
``a'' by analogy to feature A in the H$\alpha$ profile
(Figs$.$\,\ref{f3} and \ref{f4}). Feature B is the broad, red-shifted
wing around offset $\sim$\,$-$2\farcs5 and +3\farcs5, in our N93 and
S93 spectra. This feature is prominent in regions where the {\it
scattered continuum} reaches maximum intensity (see Fig$.$\,\ref{f5}
and the continuum image in Fig$.$\,\ref{f1}). Feature B is also
present in the PA3 H$\alpha$ spectra. Feature C is the slightly
red-shifted emission component observed close to the systemic velocity
from offsets +6 to +12\arcsec\ in our N93 and S93 spectra. We note
that feature C extends beyond the region where forbidden line emission
is observed (i.e., beyond the shocked lobes). From comparison with the
H$\alpha$ direct images of the nebula we conclude that feature C
originates in the weak `halo' which surrounds the bright lobes
(Fig$.$\,\ref{f1}). This halo is visible towards the east up to
offsets +20\arcsec\ in our deep, H$\alpha$ ground-based image. Feature
C is also present along slit PA3, which crosses the lobes and the halo
at the outermost parts of the east lobe.  We have labeled ``C?" the
weak, red-shifted emission beyond the west lobe in slits N93 and S93
(Fig$.$\,\ref{f4}). This tentative feature (also marginally present in
the H$\beta$ profile) has a redshift slightly larger than that of
feature C.

For slit S93, the maxima in the H$\alpha$ and forbidden (e.g$.$
[\ion{O}{1}]) line emission are observed at different spatial and
spectral positions: the H$\alpha$ emission peak is red-shifted and
located at offset $\sim$\,3\farcs5, whereas the maximum emission from
most forbidden lines is clearly blue-shifted and located at offset
$\sim$\,4\farcs2. The reason for this difference is that the H$\alpha$
emission peak is most likely the result of the superposition of features C,
B, and maybe A on top of the lobe emission and not only emission 
produced locally in the east lobe.



In Fig$.$\,\ref{f5} we present the long-slit spectra obtained along
the nebular axis (N93) in the wavelength range [4558-5331]\,\AA. The
H$\beta$ profile is similar to that of H$\alpha$, except that feature
A is less prominent for the former very likely due to a larger
extinction towards the nebular center. Features B
and C are clearly detected in the H$\beta$ profile up to (projected)
velocities of $\sim$\,290\,\kms\ (LSR) and around
\vsys, respectively.

The profile of [\ion{O}{3}]$\lambda$5007\,\AA\ is similar to that
of H$\beta$: note the maximum emission slightly red-shifted at the
position where the continuum is detected in our spectra (coincident
with feature B in H$\alpha$) and the weak emission closer to the
nebular center (coincident with feature A). The profile of the
[\ion{N}{1}] doublet (partially resolved) is different from H$\beta$ and
[\ion{O}{3}] but similar to most forbidden lines in the red.

We have detected a number of weak emission lines in this wavelength
range, most of them from Fe ions. The [\ion{Fe}{3}] lines, similarly
to the [\ion{S}{3}]$\lambda6312$ line, are observed in the region
where the scattered continuum reaches its maximum intensity and in the
inner region where feature A arises, with no trace of emission from
the outer parts of the shocked lobes. The [\ion{Fe}{2}] lines are
remarkably intense at the tips of the lobes and
show no sign of H$\alpha$-analog features, i.e$.$ they show no
emission from region A' or the region of maximal continuum emission.

At 4650\,\AA, we also detect a faint and broad emission feature, which
is only visible in the region where the continuum is observed for the
east lobe. We identify this feature with the Wolf-Rayet bump at
4650\,\AA\ (see below).

\subsection{Line identification and flux measurements}
\label{fluxes}


We list in Table \ref{table1} all the lines that we have detected in
the optical spectra of CRL\,618 together with their air wavelengths
and fluxes (undereddened). The fluxes have been estimated from slit
N93 (for other slits only the lines in the red wavelength range were
observed), integrating the emission of the pixels above the 3$\sigma$
level along the slit and in the spectral direction.

Most of the lines in Table \ref{table1} were already detected and identified in
this source from previous optical spectroscopy (see references in
\S\,\ref{intro}), however we find some new features
in this work. The most relevant of the new identifications is the
Wolf-Rayet bump (WR-bump) at $\sim$\,4650\AA\ (Fig$.$\,\ref{f5}). This
bump is a blend of high excitation lines such as
\ion{N}{3}$\lambda$\,4634-41\,\AA, and
\ion{C}{3}$\lambda$\,4647-51\,\AA\ \cite[e.g.][]{tyl93}.
The \ion{C}{4}$\lambda$4658.3\,\AA\ line, usually considered as a part
of the WR-bump, is listed separately in Table \ref{table1} together
with the blended [\ion{Fe}{3}]$\lambda$4658.1\,\AA\ line, because it
is spectrally resolved from the rest of the bump in our spectra. The
WR-bump\footnote{The
\ion{C}{3}/\ion{C}{4} lines in the WR-bump are observed in the majority of 
[WC]-type central stars of PNe \citep{leu96} but also in some normal PNe
(i.e$.$ with no [WC]-type central stars; Aller \& Keyes 1987). In the
latter, where \ion{N}{3} lines are also found, these high-excitation
transitions are the result of nebular emission. Therefore, the
presence of the WR-bump in the spectrum of \crl, while potentially
interesting and deserving further investigation, is not sufficient to
conclude that it has a [WC]-type central star.} is detected in the
east lobe, only in the region where the scattered continuum (and
feature B) is observed.  The one-dimensional blue spectrum for this
region is shown in Fig$.$\,\ref{f5}, where the presence of
\ion{He}{2}\,$\lambda$4685.55\,\AA\ (also reported by the first 
time in this work) and other high excitation (Fe?) lines, can be
noticed.

Emission around 5007\,\AA\ was detected previously with a low
signal-to-noise ratio and attributed to a blend of
$[$\ion{Fe}{2}$]$ and $[$\ion{O}{3}$]$ lines
\citep{sch81,kel92}. We have also detected 
the $[$\ion{O}{3}$]\lambda$4959 line in our deep spectra with a total
flux $\sim$\,1/3 of the $[$\ion{O}{3}$]\lambda$5007 (Table
\ref{table1}) as theoretically expected regardless of the nebular
conditions. Accordingly, the flux measured at 5007\AA\ must be 
mostly due to the $[$\ion{O}{3}$]$$\lambda$5007\AA\ line.

We have compared the absolute fluxes measured in the present work with
the fluxes measured by \citet{kel92}. For most of the lines we find
good agreement, our fluxes being on average $\sim$\,2.2 and
$\sim$\,1.6 (for the east and west lobes, respectively) times smaller
than those measured by Kelly et al$.$, as expected from the different
slit widths used in both observations (they use a 2\farcs5-wide slit,
covering the nebula almost totally in the equatorial direction and
totally along the axis). There is however a remarkable difference in
our fluxes for the $[$\ion{O}{3}$]\lambda\lambda$4959,5007\AA\ and
$[$\ion{O}{1}$]\lambda\lambda$6300,6363\AA\ lines and those measured
by Kelly et al. For the [\ion{O}{3}] transitions, these authors found
a total flux for the east (west) lobe a factor $\sim$\,0.8 (1) smaller
than ours, without correcting the different slit width. Considering
that our slit is a factor $\sim$\,2.5 smaller than theirs and the size
of the nebula (2\farcs5-3\arcsec\ in the direction perpendicular to
the slit), then we must conclude that the intensity of the
$[$\ion{O}{3}$]$ lines is now a factor 2-3 larger than 9 years
ago. \citet{tra93} also found upper limits for the flux of the
[\ion{O}{3}] lines that are also much smaller than our measurements in
spite of their broader slit (2\arcsec). We also derive a
[\ion{O}{3}]/H$\beta$ line ratio (less dependent on the slit width) a
factor $\sim$2 larger than that obtained by Kelly et al$.$ in
agreement with a recent increase of the [\ion{O}{3}] intensity.

\begin{figure}   
\epsscale{0.5}
\rotatebox{270}{\plotone{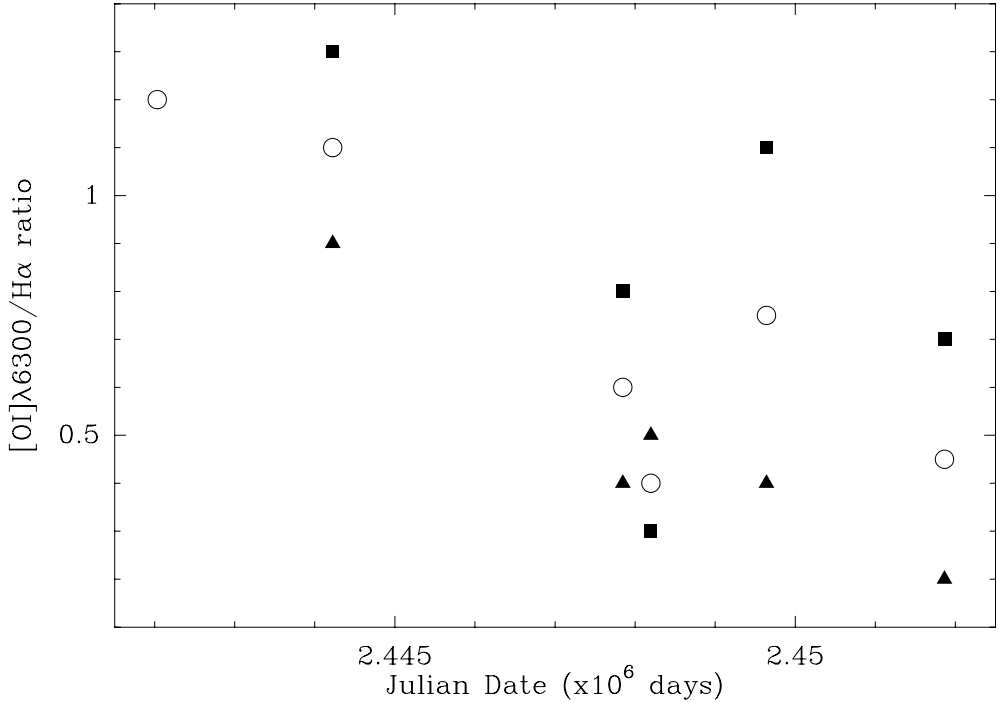}} 
\vspace{1cm}
\caption{[\ion{O}{1}]$\lambda$6300/H$\alpha$ 
line ratio measured in \crl\ by different authors (see Table
\ref{oivstime}). Filled triangles (squares) are used for the east
(west) lobe, and open circles for average values in the whole nebula.}
\label{f6}
\end{figure}

For the [\ion{O}{1}] transitions, Kelly et al$.$ measure fluxes a
factor $\sim$\,3.5 and $\sim$\,2 (for the east and west lobe) larger
than ours. Correcting by the different slit width, the intensity of
the $[$\ion{O}{1}$]$ lines has decreased by a factor $\sim$\,0.6 and
0.8 since 1990. The values of the [\ion{O}{1}]/H$\alpha$ line ratio
measured in previous works, summarized in Table \ref{oivstime} and 
Fig$.$\,\ref{f6},  confirm the progressive weakening with time
of the [\ion{O}{1}] transitions. 

\subsection{Continuum} 
\label{cont}
A relatively weak continuum is clearly visible in the blue (R900V)
long-slit spectra along the nebular axis around offset 3\arcsec\ (on
the bright east lobe, see Fig$.$\ref{f5}, top). This position
corresponds to the relative maximum in the continuum brightness
distribution seen in the direct image (Fig$.$\,\ref{f1}). We also
detect continuum emission at the same position in the red spectra
(R1200Y) for slits N93 and S93, and also, at the spatial origin for
slit PA3, but with smaller signal-to-noise ratio. The red continuum is
only visible in substantially smoothed spectra (not shown here). The
average surface brightness of the red (blue) continuum, measured from
our long-slit spectra in a 1\arcsec$\times$1\arcsec\ region around
offset 3\arcsec\ is $\sim$\,8$\times$10$^{-17}$
($\sim$\,5$\times$10$^{-17}$) \bri.
The continuum brightness has been measured in a spectral window of
80\,\AA\ in the vicinity of H$\alpha$ and H$\beta$, respectively. The
intensity and spatial distribution of the continuum obtained from our
long-slit spectra are consistent with the continuum image
(Fig$.$\,\ref{f1}).


Our measurements of the red continuum flux are in reasonably agreement
with previous estimates by \citet{kel92}. Nevertheless, the continuum
at blue wavelengths in our spectra is only a factor $\sim$\,1/1.6
weaker than the red, whereas from the data given by the previous authors we
measure a blue-to-red continuum ratio of $\sim$\,1/2.4. Note that
although the absolute flux calibration can be uncertain up to a factor
$\sim$\,50\%, the relative flux value is more accurate. This
difference is consistent with a blue brightening of a factor
$\sim$\,1.5 ($\sim$\,0.44 mag) in the last 10 years (assuming that the
red continuum flux has not varied). \citet{got76} reported an increase
of $\sim$\,2 mag in $B$-band, from measurements between 1922 and
1975. These authors derived a $B$-band brightening rate of
0.06\,mag/yr, which is roughly consistent with our results. These
authors attributed the increase of the blue continuum flux to the
expansion of the circumstellar envelope of \crl\ and the subsequent
decrease of the extinction.



\section{Spectral components as tracers of different nebular regions}
\label{sect5}

\subsection{Scattered and local emission components}
\label{features}

We find remarkable differences between the
long-slit profiles of the recombination and forbidden emission lines
in \crl. Figs$.$\,\ref{f2} and \ref{f4} clearly show the presence
of three `extra' emission features (A, B, and C) in the H$\alpha$
spectrum compared to the profile of most forbidden lines. 
Some of these features are also present in
certain forbidden lines (\S\,\ref{splines}). Here we analyze the
origin of these features.

{\it Feature A}, the intense H$\alpha$ emission observed close to the
nebular center in slits N93 and S93, {\it arises in the inner region
A'} (Fig$.$\,\ref{f1}).
This region is very bright in the $HST$ H$\alpha$ image but much
weaker in the light of [\ion{O}{1}] and [\ion{S}{2}]
\citep{tra00}. This behavior is consistent with the absence of feature
A in our spectra for these forbidden transitions. For slit S93,
feature A is blue-shifted with respect to \vsys\ by
$\sim$\,10\,\kms. The counterpart to this feature observed in other
recombination and forbidden lines is also blue-shifted: for
[\ion{N}{2}]$\lambda$6584\,\AA, feature {\it a} is blue-shifted by
$\sim$\,25 and 33\,\kms\ at position N93 and S93, respectively; for
H$\beta$, [\ion{O}{3}]$\lambda$5007\,\AA, and
[\ion{S}{3}]$\lambda$\,6312\,\AA, observed only at position N93, the
blue-shift is $\sim$8-12\,\kms. The measured Doppler blue-shift
suggests that {\it feature A is locally produced within the east
lobe}, which is approaching towards us.
For slit N93, feature A lies at \vsys, showing no apparent
shift towards the blue. We believe that in this case the spectral
position of the H$\alpha$ feature A could be altered (slightly
red-shifted) because of the presence of feature B at almost the same
spatial position, which is relatively intense and is clearly
red-shifted in the east lobe. The blue-shift of
feature A could also be intrinsically smaller for position N93 than for
S93, as suggested by the smaller blue velocities of its counterpart
(feature {\it a}) in the other lines.

{\it Features B and C} are most likely the {\it scattered component} 
of the emission in the Balmer lines for this source
(\S\,\ref{intro}). Previous spectropolarimetric observations indicate
that $\sim$\,40\% of the total H$\alpha$ flux is polarized 
\citep[with a
line polarization of about 15\%;][]{tra93}, i.e$.$ it is light
scattered by nebular dust. The polarization of the forbidden lines is,
however, much lower and this would explain the absence of these
features in these cases. Features B and C are red-shifted, although
they originate in the approaching lobe (note that the forbidden line
emission from the east lobe is blue-shifted), which clearly
points to their scattered nature: the light scattered by dust moving
away from the central source should be red-shifted for both
the approaching and receding lobe \citep[see
e.g$.$][]{sch97}. \cite{car82} also found some red-shifted H$\alpha$
emission (probably our feature B) arising in the approaching bright
lobe which they attribute to scattered emission following the same
argument.

The previous interpretation is consistent with the similar brightness
distribution of these features and the continuum 
(Figs$.$\,\ref{f1} and \,\ref{f2}), as expected if they share the 
same scattered nature. Feature B is clearly most prominent in
the region of maximum continuum emission for all the slits
positions. Moreover, features B and C are much more
intense for the east lobe than for the west one, where the scattered
continuum is much weaker and the fraction of polarized light smaller
\cite[0.5 versus 0.2, for the east and west lobe respectively;][]{sch81}.

The different profiles and spatial distributions of features B and C
in our long-slit spectra suggests that they are produced by dust in
different nebular regions. The low red-shift of feature C
($\sim$\,6-8\,\kms) and its location (beyond the forbidden-line
emitting east lobe, in the weak H$\alpha$ halo) points to dust not
located inside the lobe but beyond it, in the unshocked, slowly
expanding AGB CSE. Since the AGB CSE 
surrounds the lobes, we can expect feature C to be present
(with more or less intensity) all along the slits. This is consistent
with the presence of the diffuse halo surrounding the lobes in the
H$\alpha$ images, and not only in front of them (see below). The
tentative feature ``C?", observed at the tip of the west lobe, could be
the counterpart to feature C in the receding lobe. 

The much larger red-shift of feature B (up to LSR velocities of
$\sim$\,240\,\kms) and its location (within the lobes) is more
consistent with dust inside the shocked-lobes which is 
flowing outwards rapidly
(presumably at the same speed as the gas). We identify this
feature with the red-shifted  H$\alpha$ emission detected by
\cite{car82} in the east lobe, which they also attribute to scattered 
emission produced by fast outflowing dust. At the position where
feature B is observed, some contribution by feature C is also expected
(see above), however, the spectral resolution achieved in these
observations is not sufficient to separate both features.

The line emission that is being reflected by the dust, visible as the
halo (in the direct images) and as features B and C (in our long-slit
spectra), is unlikely to originate in the shocked-lobes themselves. In
fact, neither the halo, nor features B and C are detected in most
forbidden lines, and in particular, in the [\ion{O}{1}] line. However,
the [\ion{O}{1}] and H$\alpha$ emission {\it locally} produced in the
east lobe have comparable intensities (Fig$.$\,\ref{f4} and $HST$
images in Trammell 2000), which
should result in almost equally bright scattered components for both
lines if the reflected photons were originally produced in the
lobes. Features B and C must then result from the reflection of light
produced in the nebular center (in the central
\ion{H}{2} region or/and in region A'), where the emission of most
forbidden lines observed by us is intrinsically much less intense.

\subsection{The different nebular regions}
\label{spdist}

The presence or absence of the different spectral features
and, in general, the profiles of the different lines are diagnostic of the
physical properties (such as temperature, ionization, and electron
density) of different nebular regions in \crl, from the highly
obscured \ion{H}{2} region to the shocked lobes.

\subsubsection{The inner \ion{H}{2} region} 
This region is {\it indirectly} probed by
lines with B- or C-like features (scattered light),
namely H$\alpha$, H$\beta$, [\ion{O}{3}], \ion{He}{1}, \ion{He}{2},
[\ion{Fe}{3}], [\ion{S}{3}], and the WR-bump. The spectrum of the
central \ion{H}{2} region is thus most easily seen in the region where
the scattered continuum is observed, at offset $\sim$\,3\arcsec\ in the
east lobe.  The central \ion{H}{2} region (previously studied through
its radio-continuum emission, see \S\,\ref{intro}) is characterized by
high densities ($\gsim$\,10$^{6}$\,\cm3) and high temperatures
(13,000-15,000\,K) that provide the required conditions to produce
most of the (high-excitation) lines above (e.g., the ionization
potential of the [\ion{Fe}{3}], [\ion{S}{3}],
\ion{He}{2}, and WR-bump are 16.16, 23.33, 24.59 and $\gsim$\,30 eV,
respectively). The large critical densities for the observed
[\ion{Fe}{3}] and [\ion{S}{3}] transitions, $n_{\rm
c}$\,$\sim$\,10$^8-10^9$\,cm$^{-3}$, explain why these forbidden lines
are not de-excited by collisions in the central region of \crl.  We
note also, that the relatively large critical density of the
[\ion{O}{3}]$\lambda$5007 transition, few$\times$10$^5$\,\cm3, would
explain the weak [\ion{O}{3}]$\lambda$5007 emission from the inner
\ion{H}{2} region, which is indirectly observed at the position of maximal
continuum and feature B. Unfortunately spectropolarimetry of the lines
listed above presumably arising from the
\ion{H}{2} region has not been obtained previously, so we do not
know their polarization degree (nor the fraction of scattered flux)
and we cannot test the previous scenario. In any case it seems that
these high excitation lines are unlikely to be produced in the shocked
lobes, where the spectrum is dominated by low excitation lines.


\subsubsection{Region A'} 
This region is also very likely a relatively dense region, considering
that only forbidden lines with high critical densities ($n_{\rm
c}$\,$\gsim$\,few$\times$\,10$^4$\,\cm3) are observed therein, namely:
[\ion{N}{2}], [\ion{S}{3}], [\ion{O}{3}], and [\ion{Fe}{3}]. The
spectrum of this region (the innermost part of the east lobe, where
features A and {\it a} are observed) is remarkably different from that
of the shocked lobes but very similar to that of the central
\ion{H}{2} region. In fact, most lines are common to both regions
except for the absence of the WR-bump (which requires higher
excitation conditions) in region A' and the undetected [\ion{N}{2}]
emission (with a low critical density, $n_{\rm
c}$\,$\sim$\,2$\times$10$^4$\,\cm3) from the denser \ion{H}{2}
region. This fact suggests that region A' is more likely to be ionized
by the UV stellar radiation rather than by shocks. In this sense,
region A' represents the outermost (less dense) parts of the
\ion{H}{2} region surrounding the star. The observed relative
line intensities are not consistent with any of the different
shock-excitation models by \cite{har87} (hereafter \hrh).

The very weak emission of the [\ion{O}{1}] lines from region A' 
([\ion{O}{1}]/H$\alpha$\,$<$\,0.05) 
implies that this region is fully ionized 
\cite[e.g$.$][]{har94}. The observed [\ion{O}{1}] 
transitions have relatively high critical densities ($n_{\rm
c}$\,$\sim$2$\times$10$^6$\,\cm3) and, therefore, significant
collisional de-excitation of the involved levels is not
expected. Therefore, the weakness of the [\ion{O}{1}] transitions is
because most oxygen in region A' is ionized. Considering the similar
ionization potentials of oxygen and hydrogen, we conclude that the
ionization fraction (X=$n_{\rm e}/n_{\rm total}$) of region A', like
that of the inner \ion{H}{2} region (where [\ion{O}{1}] is not
detected) is $\sim$\,1.


\subsubsection{The shocked lobes} 
\label{lobes}
This region is probed by lines without any additional
features (A, B, or C).  The most intense of these lines ([\ion{O}{1}],
[\ion{S}{2}], and [\ion{N}{1}]) are known to originate entirely from
the gas in the lobes since for these lines, no polarization has been measured
\citep{tra93} and no red-shifted components are found in the
approaching lobe. Although no polarimetry of the rest of the lines
within this class has been made, we conclude that they must have a
similar origin given their similar profiles.

The relative intensities of the lines observed at the lobes
(dereddened using the average extinction for each lobe, see
\S\ref{sext} below) are consistent with shock-excited emission.
In particular, the flux ratios observed by us lie in between the
predicted values for bow-shock excitation models 8 and 9 by \hrh\,
with shock velocities between 50 and 100\,\kms\ and preshock density
300\,\cm3.
These range of velocities for the shocks is mainly constrained by the
detected, although weak, [\ion{O}{3}]$\lambda$5007\,\AA\ emission and
implies shock speeds slighly larger (but still comparable) to previous
estimates following similar procedures 
\citep[$V_{\rm shock}$=20-80\,\kms;][]{rie90,tra93}. 
The higher shock velocity derived by us is basically due to the higher
[\ion{O}{3}] flux obtained in our observations.

[\ion{N}{2}] and [\ion{Fe}{2}] lines show the largest intensity
contrast between the tips and the innermost regions of the lobes
amongst forbidden transitions. [\ion{Fe}{2}] lines are known to be good
tracers of astrophysical shocks (in supernovae, Herbig Haro objects,
PNe, etc): [\ion{Fe}{2}] lines are very sensitive to the high
densities and temperatures in the shocked gas; moreover, fast
shocks are able to extract substantial amounts of Fe from dust grains
\cite[e.g.][]{wel99,rei00}. Therefore, the intensity enhancement of
the [\ion{Fe}{2}] (and [\ion{N}{2}]) lines is very likely related to
the presence of intense shocks in these regions. In addition,
the strongest [\ion{Fe}{2}] line emission is expected in regions where
the compression and heating of the gas is largest, for example, at the
head of a bow shock, where the velocity component normal to the shock is
greatest. The presence of bow-shocks at the tips of the lobes of \crl,
where the maximum [\ion{Fe}{2}] emission is observed, is suggested by
bright, curved features in the direct $HST$ images of the
nebula (Fig$.$\,\ref{f1}).
Finally, the presence of intense [\ion{Fe}{2}] lines in the
shock-heated and compressed gas in the lobes of \crl\ points to
dissociative ($V_{\rm s}>$30\,\kms) J-shocks rather than C-shocks 
\citep{rei00},
consistent with the relatively large shock velocity derived from
diagnostic line ratios and the kinematics of the lobes
(see Sect$.$ \ref{str+kin}).
 
The ionization fraction in the shocked lobes is significantly 
lower than in region A' as suggested by the much larger
[\ion{O}{1}]$\lambda$6300/H$\alpha$ ratio, $\sim$\,0.5-0.9 (excluding
scattered light), in the former.  We discuss in detail the
ionization in the lobes in Sect$.$ \ref{mass}.

\subsubsection{The scattered-light halo} 
A diffuse halo surrounding the bright optical lobes of \crl\ is
visible in the H$\alpha$ direct images of the nebula
(Figs$.$\,\ref{f1}b,c). The scattered feature C in the H$\alpha$
spectra arises in this halo, and it is particularly noticeable in the
region just beyond the tip of the east lobe. 
Accordingly, the H$\alpha$ emission from this halo is very
likely light (originally arising in the \ion{H}{2} region) that
escapes preferentially in the direction of the lobes and is scattered
by the innermost parts of the AGB CSE. (The more distant regions of
the AGB CSE are visible in molecular line emission, \S\,\ref{intro}.)

\section{The inclination of the nebula}
\label{incli}

\begin{figure}    
\epsscale{0.5}
\rotatebox{270}{\plotone{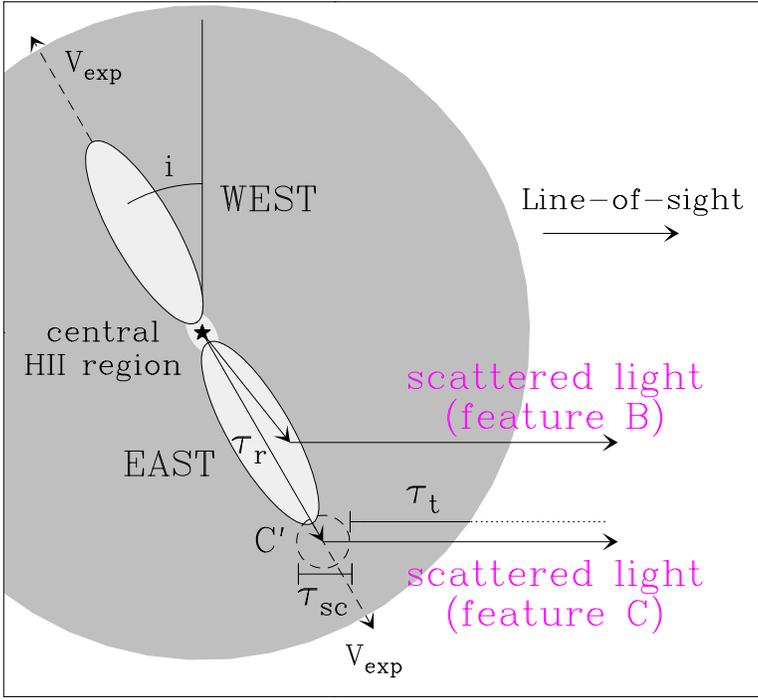}}
\caption{Schematic structure of \crl. The optical lobes and the
central \ion{H}{2} region are represented by white ellipses.  The
surrounding AGB envelope, which is radially expanding at velocity
$V_{\rm exp}$, is represented by the grey circle. The light arising in
the \ion{H}{2} region escapes preferentially through the lobes,
illuminating a compact region beyond the east lobe tip (region C')
that scatters light into the line of sight leading to feature C.  The
light from the \ion{H}{2} region is also scattered by rapidly
outflowing dust inside or in the walls of the lobes leading to feature
B. The different optical depth components ($\tau_{\rm r}$, $\tau_{\rm
t}$, \& $\tau_{\rm sc}$, see \S\,\ref{reliab}) are indicated. We do
not intend to accurately reproduce the actual shape or multiple lobe
character of \crl\ or the relative dimensions of the different nebular
components.}
\label{f7}
\end{figure}

The presence of scattered features (C and B) in the H$\alpha$ spectrum
of \crl\ offers a valuable opportunity to derive the inclination of
the nebula: by comparison of the red-shift of the scattered emission
with the intrinsic velocity of the dust (whenever these quantities can
be determined). Features B and C are produced, respectively, by: ({\sc
i}) fast-moving dust mixed with the atomic gas inside or in the walls
of the lobes; and ({\sc ii}) dust beyond the lobes, in the slowly
expanding, extended envelope which surrounds the optical lobes (see
\S\,\ref{features} and Fig$.$\,\ref{f7}). The limited spectral
resolution in our spectra does not allow us to separately measure the
red-shift of features B and C when both components are simultaneously
present (all along the east lobe up to offsets
$\sim$\,8\,\arcsec). However, at the tip of the east lobe, only
feature C is observed in the H$\alpha$ spectra, and the red-shift of
this feature can be accurately determined. By comparison of this
red-shift with the intrinsic expansion velocity of the dust in the AGB
CSE producing feature C, $V_{\rm exp}$\,$\sim$\,17.5\,\kms\ LSR
(\S\,\ref{intro}), we can readily derive the nebular inclination.



\begin{figure}    
\epsscale{1.1}
\plottwo{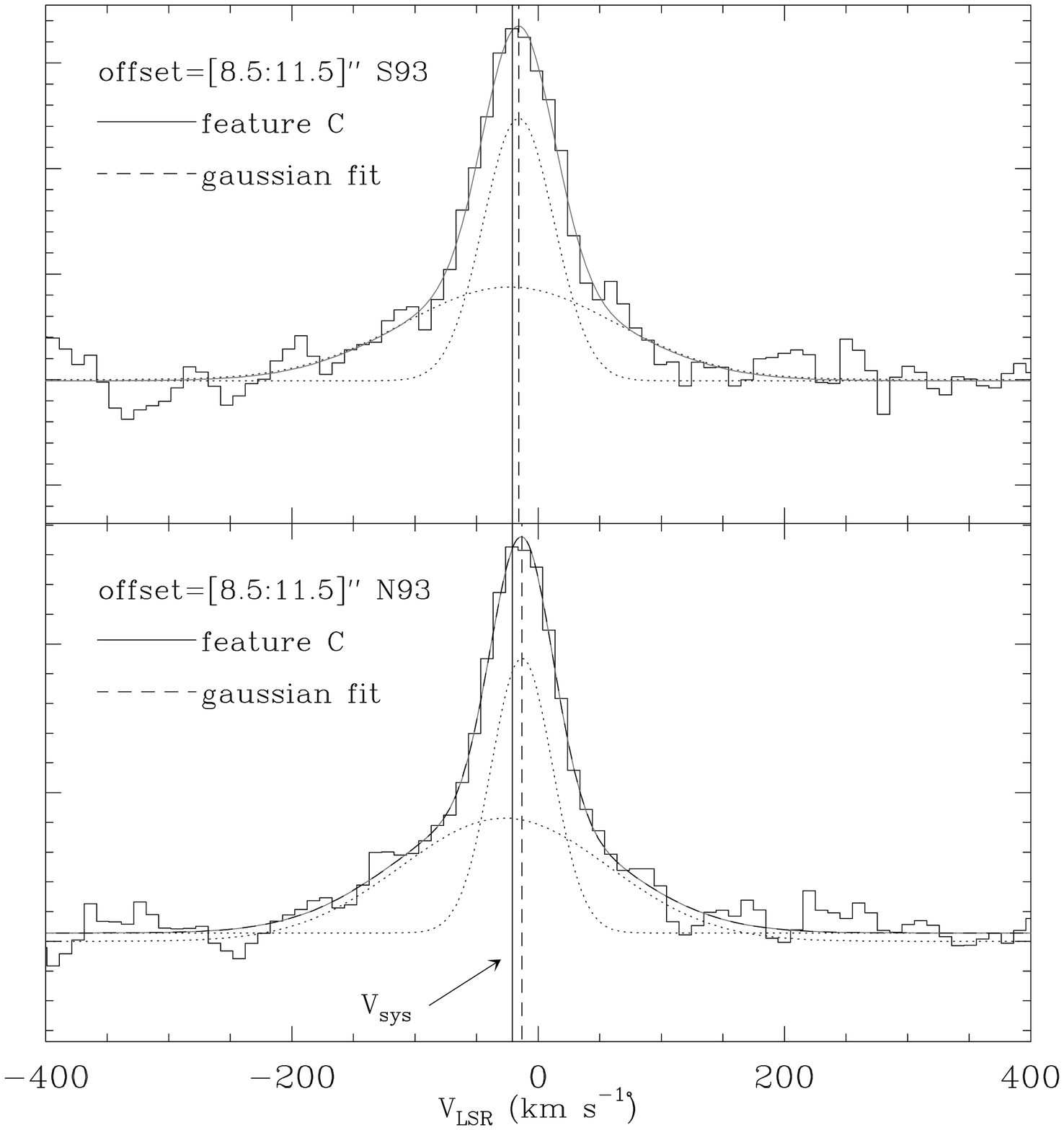}{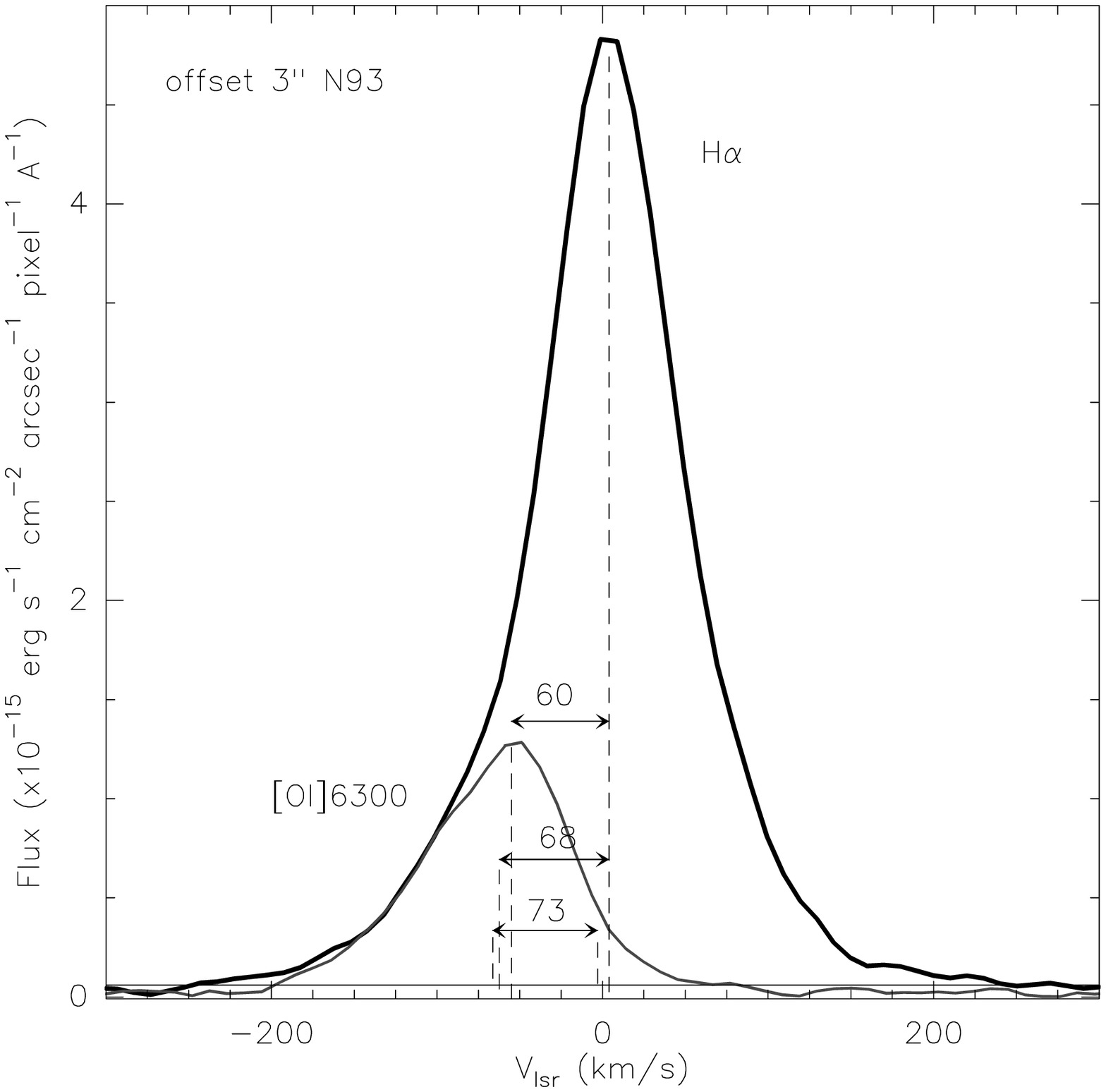}
\caption{{\it Left)} Spectral profile of feature C at the tip of the east lobe 
(solid-line histogram): the H$\alpha$ long-slit spectra have been
spatially averaged from offset 8\farcs5 to 11\farcs5 for slits
positions N93 (bottom) and S93 ({\it top}). The core of feature C is roughly
symmetric and has been approximated by a gaussian function
(dashed-line). The center of the gaussian (vertical dashed-line) is
red-shifted with respect to the systemic velocity (vertical
solid-line). The wings of the feature have been also fitted by a
gaussian function. {\it Right)} One-dimensional H$\alpha$ (dark line) and
[\ion{O}{1}]$\lambda$\,6300\,\AA\ (light line) spectra obtained at
offset 3\arcsec\ for slit N93. The red-shifted H$\alpha$ emission,
contrary to the blue-shifted [\ion{O}{1}] transition, is dominated by
features A, B, and C (see text) rather than by emission locally
produced by the shocked gas in the east lobe.}
\label{f8}
\end{figure}

The red-shift of feature C is apparent in Fig$.$\,\ref{f8} (left),
where we show its spectral profile derived from slits N93 and S93
integrating the long-slit spectra from offsets 8\farcs5 to
11\farcs5. The red-shift of feature C for both slit positions has been
obtained by fitting a Gaussian function to the core of the feature,
which is roughly symmetric.
For N93 and S93, the red-shift with
respect to \vsys\ (given by the Gaussian center) 
is 8$\pm$1.5 and 6$\pm$1.5\,\kms\ respectively. 
%
These values yield an inclination (with respect to the plane of the
sky) of $\sim$\,27\degr$\pm$6\degr\ (from slit N93) and
20\degr$\pm$5\degr\ (from slit S93) for the dust reflecting the
H$\alpha$ photons ahead of the east lobe.

As we have shown, feature C beyond the east lobe has its counterpart
in the weak halo seen in the H$\alpha$ images of the nebula at the
same position (Fig$.$\,\ref{f1}). The orientation of this halo and the
pure emission lobe in the plane of the sky is roughly the same. This
is most likely due to light from the core escaping preferentially along 
the lobes of \crl. Hence we 
can expect the previous values of the inclination to be a relatively
good representation of the mean nebular inclination (for an overall
axial symmetry).

The different inclinations ($\Delta$$i$=7\degr) derived for
feature C in slits S93 and N93 could indicate different orientations
of the different lobe components along the line-of-sight, a likely
possibility given their different orientations in the plane of the sky
($\Delta$PA$\sim$\,15\degr).

We can also derive $i$ by measuring the velocity of feature B ($V_{\rm
B}$) with respect to the local emission component ($V_{\rm local}$,
obtained from the forbidden lines) at the same position.  $V$=$V_{\rm
B}$-$V_{\rm local}$ is a direct measurement of the intrinsic expansion
velocity of the lobes, therefore $i$=sin$^{-1}(V_{\rm
local}-\vsys)/(V_{\rm B}-V_{\rm local})$. From our spectra, we can
derive only a lower limit to the red-shift of feature B because the
H$\alpha$ profile has also important contributions from features A and
C at the position where feature B is observed (offset
$\sim$3\arcsec). This leads to a combined H$\alpha$ profile with a
smaller red-shift than that of feature B alone and, therefore, only a
lower limit to the intrinsic expansion velocity of the lobes. 
The lower limit for $V$ translates into
an upper limit for $i$.

In the right panel of Fig$.$ \,\ref{f8}, we show H$\alpha$ and
[\ion{O}{1}]$\lambda$6300\,\AA\ 1D-spectra at the position where the
scattered continuum and feature B reach maximum intensity in the east
lobe for slit N93. The velocities of the H$\alpha$ and [\ion{O}{1}]
lines measured (with uncertainties of $\pm$1.5\,\kms) at the peak are
$V_{\rm B}$=5\,\kms, $V_{\rm local}$=$-$55\,\kms, which yields
$V$\,$>$\,60\,\kms\ (at this position) and $i$\,$<$34\degr. If,
instead of using the line peaks for the velocity measurements, we use
the centroids of the full width at half maximum ($V_{\rm B}$=5\,\kms,
$V_{\rm local}$=$-$63\,\kms, $V$\,$>$\,68\,\kms), or the centroids of
the full width at a 3$\sigma$ level ($V_{\rm B}$=$-$6\,\kms, $V_{\rm
local}$=$-$67\,\kms, $V$\,$>$\,73\,\kms), we get values of
$i<38$\degr\ and $i<39$\degr, respectively.

The mean value of, and upper limit to, the inclination derived above are
smaller than the inclination previously obtained by \cite{car82}
(hereafter \cs). These authors obtain $i$=45\degr, based on the
analysis of the H$\alpha$ scattered component produced by the fast
dust within the lobes.  We believe this difference can be due to
several reasons:

First, the value given by \cs\ for the intrinsic speed of the
outflow, 80\,\kms\ is really a lower limit to $V$ because of
the presence of multiple components in the H$\alpha$ profile (as
explained above).  Second, the value for the radial velocity
difference between the east and west lobes (which is equal to
2$V$sin\,$i$) measured by \cs\ from the forbidden line emission
(e.g$.$ [\ion{N}{2}]$\lambda\lambda$6548,6583,
[\ion{S}{2}]$\lambda\lambda$6716,6731), 114\,\kms, is most
likely an overestimate. This is because the radial velocity difference
between the lobes is not constant along their length (see
Figs$.$\,\ref{f2} and
\ref{f4} and Table \ref{table2}). In particular, 
the 114\,\kms\ velocity difference occurs only at the tips of the
lobes (at offsets $\pm$6\arcsec). But in the region where the most
intense H$\alpha$ scattered emission is observed (offset
$\sim$3\arcsec), coincident with the position where \cs\ measure the
red-shift of the H$\alpha$ scattered component, we observe a smaller
radial velocity difference, $\sim$\,90\,\kms. Thus, if we use a larger
value of the outflow speed ($>$80 \kms) and a smaller radial velocity
difference ($\sim$\,90\,\kms) in the
\cs\ method, we get an upper limit to the inclination 
of $<$\,34\,\degr, which is consistent with our previous estimates.

Finally, a low mean value of the inclination is in better agreement
with the similar average extinction deduced for the east and west lobe
(see \S\,\ref{sext}).

\section{Kinematics}
\label{str+kin}

\subsection{The shocked lobes}
In order to study the kinematics in the lobes of CRL\,618 
we have chosen the most intense forbidden lines, which originate in the lobes
and have profiles uncontaminated by scattered light components.  Since
the [\ion{N}{2}] lines have also some contribution from reflected
light (Trammell, Dinerstein, and Goodrich, 1993, found that the
polarization of these lines is $\sim$\,6\%), we have based most of our
analysis on the intense, unpolarized [\ion{O}{1}] and [\ion{S}{2}]
lines.


The [\ion{O}{1}] and [\ion{S}{2}] profiles along the slit N93 for the
west and east lobe are approximately point-symmetric with respect to
the spatial origin and systemic velocity (Fig$.$ \ref{f2}), suggesting
that they have probably a similar kinematical structure. This result
is not in principle expected from the different morphology of the east
and west lobes. The projected radial velocity gradient is not constant
in either lobe. In the innermost parts of the lobes the projected
velocity decreases with the distance from the nebular center, this
behavior being more pronounced in the east lobe (Table
\ref{table2}).  In the outer parts of the lobes, the velocity
increases with the distance, reaching projected velocities (with
respect to \vsys) up to 70 and 80\,\kms\ at the tip of the
east and west lobe, respectively. In these regions, the emission
arises in the compact, bow-shaped emitting knots at
$\pm$6\arcsec. Considering the low inclination for the nebula
derived in \S\,\ref{incli}, $i\sim$24\degr, the lobes of
\crl\ are expanding with velocities up to $\sim$\,200\,\kms\ (at the
tips). The changing radial velocity gradient
observed in our low angular-resolution (ground-based) spectra may
result from superposition of multiple kinematical components as
suggested by the complex morphology and small scale ($\sim$0\farcs2)
structure seen in the $HST$ images. Of course it is
also possible that there are acceleration and deceleration processes
operating in the lobes.

\begin{figure}    
\epsscale{0.5}
\plotone{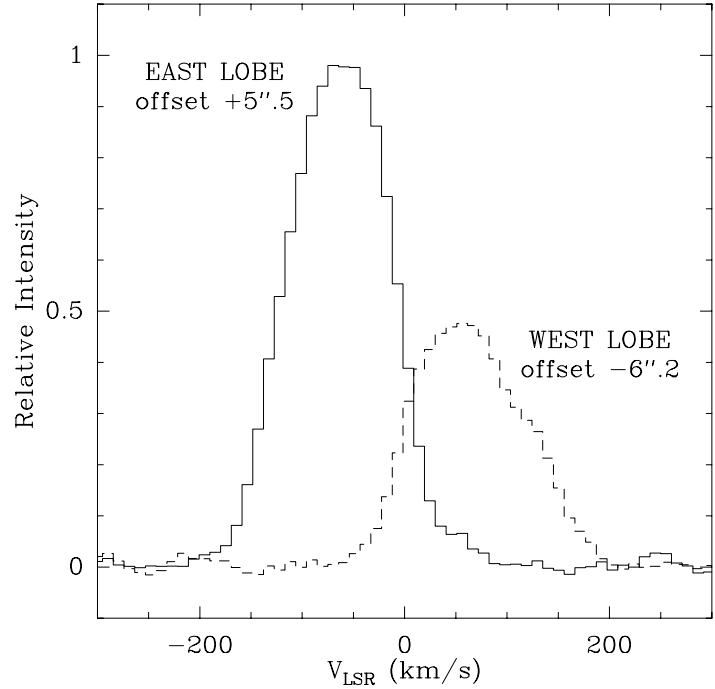}
\caption{Spectral profile of the [\ion{O}{1}]$\lambda$6300 line 
for slit N93 at the tips of the east and west lobe, coincident with
the bow-shaped emitting regions.}
\label{f9}
\end{figure}

The FWHM of the forbidden lines along the lobes
also increases outwards, from 60--75\,\kms\ closer to the nebula
center, to $\sim$\,150\,\kms\ at the lobe tips (Table
\ref{table2}). The large line widths measured at the tips of the lobes
(Fig$.$\,\ref{f9}) are consistent with the bow-shaped features being
regions accelerated by bow-shocks: the different orientations of the
velocity vectors in a bow-shock structure produce a large velocity
dispersion (see e.g$.$ \hrh). The total width (at zero
intensity level) of the line does not strongly vary along the lobes
but rather is fairly constant with a mean value of $\sim$\,200\,\kms.

The spectral profile of any low-excitation emission line from a
bow-shock can be used to estimate the shock velocity, $V_{\rm s}$. The
shock velocity equals the full width (at zero intensity) of the line
for radiating bow-shocks, independent of orientation angle, pre-shock
density, bow-shock shape, and pre-shock ionization stage (\hrh). We
derive $V_{\rm s}$=200-230\,\kms\ for the bow-shocks seen at
the tips of the east and west lobes. This value agrees well with the
value of $V_{\rm s}$ inferred directly from the projected velocities
of these regions assuming an inclination of 24\degr. This result
therefore supports the low inclination of the nebula we have derived
(\S\,\ref{incli}). In addition, the centrally-peaked profiles in
Fig$.$\,\ref{f9} are consistent with the predictions for bow-shock
models by \hrh\ with $i<$25\degr\ (for larger inclinations the model 
profiles show two well separated peaks).


The shock velocity derived from the observed line ratios and their
comparison with predictions of bow-shock excitation models by \hrh\
($V_{\rm s}$\,$<$\,90\,\kms, \S\,\ref{lobes}) is smaller than that
directly derived from the line profile. Similar differences are also
present for many Herbig-Haro objects observed and modeled by these
authors suggesting that the existing models (necessarily simplistic)
cannot accurately reproduce the complex excitation and
spatio-kinematical structure of the shocked material. Moreover, the
preshock density in the \hrh\ models, 300\,\cm3, is very likely lower
than the actual value (see \S\,\ref{discuss}). The shock velocity
obtained from the line profile is probably less affected by errors in
the model, since it arises only from geometrical considerations, and
may better represent the actual velocity of the shock. Only if the
line emission from the shocked post-AGB wind itself (which is moving
faster than the forward shocks) was significant compared to that from
the shocked AGB ambient material (unlikely to be the case in \crl;
C-F$.$ Lee and Sahai, in preparation), then the line FWZI would be an
upper limit to $V_{\rm s}$. $V_{\rm s}$ derived from the line ratios
depends on poorly known parameters like, e.g., the presence and
strength of magnetic fields. Models with higher magnetic fields
require substantially larger shock velocities to reproduce the same
line ratios \citep{har94}. For example, for a pre-shock density
$\gsim$10$^5$\,\cm3 (\S\,\ref{discuss}), magnetic fields of
$\sim$\,30$\mu$G are required to reproduce the dereddened
[\ion{N}{2}]$\lambda$6583/[\ion{O}{1}]$\lambda$\,6300 ratio
($\sim$\,0.5, see Table \ref{table1} and \S\,\ref{sext}) for a planar
shock with speed $V_{\rm s}$\,=\,90\,\kms. To produce the same ratio
when $V_{\rm s}$\,=\,200\,\kms, the required magnetic field is much
larger ($>$3000$\mu$G).



Finally, from a comparison of the spectra obtained for slits N93, S93,
and PA3, we infer some differences in the kinematics
of the different lobe components (note that each lobe of CRL\,618
consists of at least two well separated features,
Fig$.$\,\ref{f2}). For example, the south sub-component of the east
lobe (better probed by slit S93) is expanding 5-20\,\kms\
faster than the north one. Similarly, the northern sub-feature of the
west lobe (better seen in slit N93) is 5--20\,\kms\ faster
than the southern. These differences could be due to different 
absolute ejection velocities or inclinations of the lobe
components.  We note also that, in general, N93 and S93 provide
similar FWHM for the west lobe but for the east lobe, the largest FWHM
are measured from slit S93, where the line FWHM reaches up to
$\sim$\,180\,\kms. The larger widths in the southeast lobe
sub-component are unlikely to be related to projection effects and must be
due to intrinsic differences in the kinematics.

\subsection{Region A' and the inner \ion{H}{2} region}

From the analysis of the H$\alpha$ profile we can also obtain valuable
information on the kinematics in regions of the CRL\,618 nebula other
than the shocked lobes. The presence of the blue-shifted, intense
H$\alpha$ emission component in the east lobe close to the nebula
center (feature A, Fig$.$\,\ref{f2}) indicate ionized gas moving
away at moderate velocity (15-30\,\kms) from the central star. We
have measured the velocity of this feature from the [\ion{N}{2}]
lines, which are less contaminated by scattered light (note that the
blue-shift of feature A in the H$\alpha$ spectra is masked by
superposition with the intense, red-shifted feature B and, probably also, 
C).

The kinematics of the \ion{H}{2} region cannot be so straightforwardly
obtained from our data, since the emission from this region is seen
only after being reflected by the dust (through features B and C) and
therefore the resulting line profile is affected by the distribution
and kinematics of the dust.  The FWHM of feature C, observed
at the tip of the lobes (Fig$.$\,\ref{f8}), and deconvolved with the 
spectral resolution of our data is $\sim$\,60\,\kms,
suggesting an expansion velocity for the gas in the
\ion{H}{2} region of $\sim$\,30\,\kms. 
This figure has to be considered as an upper limit, 
since the reflecting dust could show a
range of inclinations resulting from the opening angle of the lobes
(up to 15\,\degr\ according to the lobe aperture 
in the plane of the sky). Our result
is consistent with the low expansion velocity of the \ion{H}{2} region
measured from the profile of H radio recombination lines
\citep[20\,\kms;][]{mar88}. The weak wings of feature B in the
H$\alpha$ and H$\beta$ lines extend up to LSR velocity
$\sim$\,240\,\kms\ (at a level of 5$\sigma$). Such extended red-wings
are most likely the result of the large velocities of the reflecting grains
within the lobe (up to $\sim$\,200\,\kms) and also partially to the
lobe opening angle, however, the presence of rapidly outflowing material 
in the \ion{H}{2} region cannot be ruled out.

\section{Extinction}
\label{sext}


We have estimated the relative variation of the extinction along the
lobes of CRL\,618 by comparing the H$\alpha$ and H$\beta$ spatial
profiles along the axis (from slit position N93). The two long-slit
spectra have been aligned in the direction of the slit using the west
lobe brightest knot (at the tip of the lobe) as our spatial
reference. We estimate that errors in the alignment are $\lsim$\,1
pixel ($\lsim$0\farcs2). The uncertainty in the positions of the slits
used for obtaining the H$\alpha$ and H$\beta$ spectra in the direction
perpendicular to the slit is a few pixels and therefore,
smaller than the seeing (1\farcs6, \S\,\ref{obsop}). The
spatial distribution for both transitions and the adjacent
continuum was obtained integrating our long-slit spectra in the
spectral direction. In order to derive the Balmer decrement, the
continuum emission was first subtracted from the H$\beta$ spatial
profile but not from the much brighter H$\alpha$ emission (in this
case, the continuum contribution is negligible, $\lsim$1\%).

The spatial distribution of the extinction derived from the
H$\alpha$/H$\beta$ ratio is shown in Fig$.$\,\ref{f10} (top panel). In
this figure, the optical depth at 4861.3\,\AA, $\tau_{4861}$, has been
calculated assuming the Galactic reddening curve
by \citet{how83} and an intrinsic Balmer decrement of 3. This value of
the Balmer decrement allows easy comparison of our results with
previous estimates of the extinction, which usually assume
H$\alpha$/H$\beta$=3. Moreover, the previous value
is expected for high-velocity shocks like those in
\crl\ (\S\,\ref{str+kin}), for 
which the intrinsic Balmer ratio approaches the recombination value
\citep[collisional excitation increases the intrinsic Balmer ratio only 
for low-velocity shocks, $<$\,70\,\kms;][]{har94}. The conversion from
$\tau_{4861}$ to extinction in magnitudes in the $V$ band,
$\tau_{4861}$=0.93$A_V$, has been derived using the extinction law
parametrization by \citet{car89} and assuming the ratio of total-to-selective 
absorption, $R_V$, equal to 3.1. The error-bars in
Fig$.$\,\ref{f10} are statistical, and do not account for systematic
errors (e.g. arising from an incorrect value of the intrinsic Balmer
decrement, absolute flux calibration or small misalignments between
the H$\alpha$ and H$\beta$ spectra).

The H$\alpha$/H$\beta$ ratio clearly varies along the nebular axis
(Fig$.$\,\ref{f10}). This variation is most likely not due to a change
of the intrinsic Balmer ratio, since we do not expect the shock
characteristics (e.g$.$ shock speed and pre-shock density) to vary
sufficiently along the nebular axis. Neither the velocity of the
shocks nor the electron density show systematic variations along the
lobes (see \S\,\ref{str+kin} and \S\,\ref{physical}). Accordingly, the
observed variation of the H$\alpha$/H$\beta$ ratio must be related to
extinction by nebular dust.

The nebular extinction varies along the axis of \crl\ as follows: 
the maximum extinction is produced close to the nebular center and then
gradually decreases towards the outermost parts of the lobes. 
The extinction in the innermost nebular regions is particularly uncertain
and is very likely underestimated by a large factor. 
In fact, from the $HST$ images we can see that
the emission from the 2\arcsec-region around the nebular center
is well below the noise, suggesting a very high obscuration there 
(see Fig$.$\,\ref{f1}). We deduce that the extinction towards the
central star of \crl\ is $A_V$\,$>$\,10 magnitudes, given the upper limit
to the stellar continuum level estimated from the R1200Y spectra,
$\lsim$\,2.5$\times$\,10$^{-16}$\flux, and the expected flux from a
30,000\,K black body, 2.5$\times$\,10$^{-12}$\flux.  Our
low-resolution, ground-based H$\alpha$ and H$\beta$ spectra, used to
calculate the extinction, show some weak emission from the innermost
nebular region. However, this is a result of the poor spatial
resolution in these spectra, which spreads out some light from nearby
nebular regions in the lobes (where the extinction is smaller) into
the nebular center (with, very likely, much larger extinction). The result 
is an underestimate of the extinction in these inner regions as deduced
from the observed Balmer decrement.

Beyond the central region, the
extinction decreases with the distance from the nebular center reaching a
relative minimum around offsets $\sim$\,$\pm$6\farcs5. 
At offsets larger than $\pm$7\arcsec\ (beyond the tips of the lobes) 
the extinction increases again.

We note that the extinction curve along the nebula is quite symmetric,
i.e$.$ the extinction is similar for the east and west lobe: the weak
west lobe appears on average only about 15\% more extinguished than
the bright east lobe. The difference is even smaller at the tip of the
lobes. This result indicates that the lower H$\alpha$ surface
brightness for the west lobe compared to the east (by a factor
$\sim$\,5) is intrinsic and not due (at least totally) to a much
larger extinction of the receding lobe. In fact, most of the H$\alpha$
emission from the east lobe arises from region A', which does not have
a counterpart in the west lobe. The similar extinction for the east
and west lobe is consistent with the similar brightness of both lobes
for most forbidden lines (e.g$.$ [\ion{O}{1}], [\ion{S}{2}], etc), for
which the east lobe is only a factor $\sim$\,1.5 brighter than the
west lobe.  Finally, the similar extinctions derived for the east and
west lobe suggest a low inclination of the nebular axis with respect
to the plane of the sky (assuming that most of the extinction is
produced by the AGB CSE surrounding the lobes, \S\,\ref{intro}),
consistent with the value we derive from the analysis of the scattered
H$\alpha$ emission ($i$$\sim$24\degr, \S\,\ref{incli}).

Our results on the extinction in \crl\ are in good agreement with
previous estimates using the Balmer decrement method: 
\cite{kel92} derive an average extinction in the nebula of
$A_{\rm V}$=4.5$\pm$0.2, which is within the nebular extinction range
derived by us (Fig$.$\,\ref{f10}). From the analysis of
[\ion{S}{2}]$\lambda\lambda$6716,6731 and
[\ion{S}{2}]$\lambda\lambda$4069,4076 lines, these authors also found
that both the east and west lobe have similar extinction values,
$A_{\rm V}$=2.1$\pm$0.2\,mag and $A_{\rm V}$=2.9$\pm$0.3\,mag,
respectively.

\subsection{Effect of the scattering on $A_V$: radial extinction 
beyond the lobes}
\label{reliab}

The estimate of the absolute extinction along the line of sight from
the Balmer decrement in an object like CRL\,618 is problematic mainly
because of the contribution of scattered light to the total flux of
the \ion{H}{1} recombination lines, as already noticed by, e.g.,
\citet{tra93,kel92}.  The effects of such contribution on the 
value of the observed H$\alpha$/H$\beta$ ratio are difficult to
quantify: on the one hand, the scattering will produce an `artificial'
decrease of the observed H$\alpha$/H$\beta$ ratio due to the highest
reflection efficiency for blue photons; on the other hand, the
extinction of the light from the innermost regions, where it
originates, to the regions where it is scattered by the dust (radial
extinction) will produce reddening which may or may not compensate the
previous artificial increase of the H$\alpha$/H$\beta$ ratio. 

The presence of nebular dust is expected to affect not only the
absolute value of the extinction but also its relative variation from
one region to another in the nebula.  In regions where both local and
scattered H$\alpha$ and H$\beta$ emission are present, the optical
depth derived from the Balmer decrement, $\tau$, is not only the
line-of-sight or tangential optical thickness, $\tau_t$ (from the
lobes to the observer), as desired, but a complex function of
$\tau_t$, the radial optical depth, $\tau_r$ (from the star up to the
lobes where the light is being reflected), and the relative intensity
of the Balmer emission locally produced within the lobes and in the
inner \ion{H}{2} region. All these quantities will vary
along the nebula in a different way and they will combine to yield the
extinction profile in Fig$.$\,\ref{f10}.

At the tip of the \crl\ east lobe (and tentatively in the tip of the
west lobe) only H$\alpha$ emission from feature C (and ``C?'') is
observed, i.e$.$ the Balmer emission is dominated by the scattered
component (Fig$.$\,\ref{f4}). In a region like this, where the local
emission is negligible with respect to the scattered component, 
$\tau$ only depends on $\tau_r$ and $\tau_t$ in a relatively simple way:

\begin{equation}
\tau = \tau_t + \tau_r - {\rm ln} \tau_{\rm sc}
\end{equation}

\noindent
where $\tau_{\rm sc}$ is the scattering optical thickness of the small
region at the tip of the east lobe that produces feature C (region C';
see Fig$.$\,\ref{f7}) and the optical depths $\tau_r$ and $\tau_t$
include both scattering and absorption. This equation is valid if the
scattering and aborption optical depths in region C' are small
($\ll$\,1), as expected beyond the tips of the lobes. In this case,
$\tau_{\rm t}$ is mostly due to foreground dust (not directly
illuminated by the central source) between region C' and the
Earth. Using this equation, we analyze the spatial variation of the
extinction in Fig$.$\,\ref{f10} to study the structure of the
lobes. At the tips of, and beyond, the lobes (offsets
$\gsim$\,7\arcsec), the extinction increases with the distance from
the nebular center (Fig$.$\,\ref{f10}). This increase, which is not
expected a priori given the progressive decrease of $\tau$ with
distance for inner regions, can be explained if $\tau \approx \tau_r$
in this region, i.e., if the radial extinction is larger than the
tangential component. In fact, the tangential optical depth will
decrease (not increase) with the distance from the central star in the
plane of the sky, $p$, for any reasonable density law for the outer
circumstellar envelope (e.g$.$ $\tau_t \propto p^{-1}$ for a density
law $\rho=\rho_0 (r_0/r)^2$). However, the radial extinction is
expected to increase with $p$, as observed, for any reasonable density
law within the lobes.

We have computed the radial extinction as a function of the observed
spatial offset, $p$=$r$\,${\rm cos}i$, for two different density laws
within the lobes of \crl: ({\sc i}) $\rho=\rho_0(r_0/r)^2$, which
leads to $\tau_r \propto (1/p_0-1/p)$; and ({\sc ii}) $\rho =
constant$, which yields $\tau_r \propto p-p_0$. The increasing
extinction beyond offset +7\arcsec, can be reproduced for both laws
(note that, in fact, for $p \sim p_0$, the radial extinction $\tau_r$
in case {\sc i} can be approximated by a straight line, which
corresponds to case {\sc ii}) but always requires values of $p_0
\gsim$\,7\arcsec. (We will not discuss case {\sc ii} further because
$\rho \propto r^{-2}$ best represents the density in most AGB
circumstellar envelopes.) A small $p_0$ ($\ll$7\arcsec) results in the
extinction reaching a constant value ($\tau_r \propto 1/p_0$) very
quickly (Fig$.$\,\ref{f10}), which is inconsistent with the data.

The large $p_0$ obtained implies that there is a substantial component
of the radial extinction which is produced by dust {\it outside} of
the lobes (i.e., in the AGB CSE). We can only set an upper limit to
the component of the radial extinction produced by dust {\it inside}
the lobes (presumably filled by the post-AGB wind), based on the
minimum value of $\tau_{4861}$ (about 2) obtained at the tips of the
lobes (offsets of $\pm$6\farcs5\arcsec). Both the above components
(referred to as $\tau_{\rm AGB}$ and $\tau_{\rm pAGB}$)
and their sum are shown in Fig$.$\ref{f10}.

\subsection{The density contrast between the AGB and post-AGB wind}
\label{massloss}

Assuming that the gas filling the lobes is the post-AGB (pAGB) wind, 
we now derive $\rho_{\rm pAGB}$/$\rho_{\rm AGB}$ by comparing
$\tau_{\rm AGB}$ and $\tau_{\rm pAGB}$.  For a
stationary outflow the mass loss rate per solid angle is 
\mloss$_{\Omega} \equiv {\rm d}\mloss/{\rm d}
\Omega$=$r^2$$\rho$$V$, and the corresponding radial optical thickness is: 

\begin{equation}
\tau_r \propto \int_{r_0}^{r} n_{\rm dust}{\rm d}r = \kappa \frac{\mloss_{\Omega}}{V} (\frac{1}{r_0}-\frac{1}{r}) 
\label{opa}
\end{equation} 

\noindent
where $r$ is the distance from the star, $r_0$ is the inner radius,
$\rho$ is the gas mass density, $V$ is the wind expansion
velocity, and $\kappa$ depends on the grain properties and the
gas-to-dust mass ratio. Combining Eq$.$\,\ref{opa} and the expression
for \mloss$_{\Omega}$ above, we derive that:

\begin{equation}
\rho \propto \frac{\mloss_{\Omega}}{r^2V} \propto \frac{\tau_r}{r^2(\frac{1}{r_0}-\frac{1}{r})}
\label{rhoeq}
\end{equation} 

From our fitting to the radial optical depths (\S\,\ref{reliab}),
which gives for the AGB CSE inner radius, $r_{\rm
0,AGB}$\,$\gsim$\,7\farcs5, and adopting for the post-AGB wind inner
radius, $r_{\rm 0,pAGB}$\,$\sim$0\farcs06 \citep[from the
circumstellar dust model by][]{kna93}, we obtain:

\begin{equation}
\frac{\rho_{\rm AGB}}{\rho_{\rm pAGB}} 
\gsim 600(\frac{0\farcs06}{r_{\rm 0,pAGB}}) 
\frac{\kappa_{\rm pAGB}}{\kappa_{\rm AGB}}
\label{densratio2}
\end{equation}

\noindent 
where the lower limit arises from the upper limit to $\tau_{\rm
pAGB}$. The density ratio above depends on the poorly known inner
radius of the post-AGB wind as well as on the unknown but expected
differences in the dust-to-gas ratio and dust grain properties in the
AGB CSE and in the fast post-AGB wind. Moreover, the intrinsic
$\rho_{\rm AGB}/\rho_{\rm pAGB}$ ratio is likely to be lower because
the region beyond the lobe tips is overdense compared to the original,
unshocked, AGB CSE (since it contains material swept-up by the
post-AGB wind). The AGB-to-pAGB density ratio would also be
significantly lower if the gas-to-dust ratio in the post-AGB wind was
much higher than that in the AGB CSE, perhaps as a result of
inefficient grain formation in the post-AGB wind emanating from the
vicinity of a 30,000\,K star.

Finally, as argued earlier, for offsets $<$7\arcsec\ the radial
and tangential optical depth cannot be separately quantified from
$\tau_{4861}$. The smooth decrease of $\tau_{4861}$ with distance from
the central star may suggest that the derived extinction is dominated
by the tangential component (as shown above, the radial component
would increase or remain constant with the distance from the star). This
is, in particular, most likely true for the west lobe, for which the
fraction of scattered light in the H$\alpha$ and H$\beta$ emission
lines is small. The similarity between
the extinction profile for the east and west lobe as well as the
expected lower extinction derived for the east (approaching) lobe,
again suggest that $\tau_{4861}$ is probably dominated by the
tangential component. However, the derived extinction, which may still
be affected by the scattering, does not allow us to perform a
detailed modelling of $\tau_t$ to study the outer AGB CSE.

\section{Density and nebular mass}
\label{physical}

\subsection{The electron density distribution}
\label{ne}

We have used the [\ion{S}{2}]$\lambda$6716/$\lambda$6731 doublet ratio
to estimate the electron density, $n_{\rm e}$, along the shocked lobes
of \crl. The spatial profiles along the nebula
axis (slit N93) for the two doublet lines and their ratio are plotted
in Fig$.$\,\ref{f10} (middle panel). We only show the ratios wherever
the line intensities are over 3$\sigma$.
There are no alignment errors to contend with since the two lines of
the doublet were observed simultaneously.


Within the observational errors, the
[\ion{S}{2}]$\lambda$6716/$\lambda$6731 ratio does not present a
systematic variation along the east or west lobes, with a mean
values of 0.55 and 0.50, respectively. The same average values
are obtained for slit S93. Considering the
[\ion{S}{2}]$\lambda$6716/$\lambda$6731 ratio versus electron
temperature and density contour plots by
\cite{can80} and an average electron temperature 
$T_{\rm e}$\,=\,10,000\,K (\S\,\ref{intro}), we derive densities
for the east (west) lobe of $\sim$\,[5-8]$\times$10$^3$
($\sim$\,8$\times$10$^3$-10$^4$)\,\cm3. 
These figures are close to the critical densities for the
[\ion{S}{2}]$\lambda$$\lambda$6716,6731\,\AA\ transitions, however,
the good agreement between our results and previous estimates using
different diagnostic lines
\citep{kel92}, suggest that collisional de-excitation of the
[\ion{S}{2}] lines is not critical, at least, for most regions in the
lobes. 

There are some non-systematic variations of the [\ion{S}{2}] doublet
ratio within the lobes which seem to be larger than their
corresponding error-bars. These variations could suggest density
inhomogeneities along the nebula (from $\sim$\,2,000\,\cm3\ to
$\sim$\,2$\times$10$^4$\,\cm3).

\subsection{Atomic and ionized mass and ionization fraction}
\label{mass}

We have estimated the total mass of atomic and ionized gas, $M_{\rm
H}$ and $M_{{\rm H}^+}$, in the lobes of \crl\ using the mean electron
densities derived above and the total energy radiated by
[\ion{O}{1}](6300+6363) and H$\alpha$, respectively.  The
[\ion{O}{1}] (H$\alpha$) intensity is proportional to the product of the
electron density and the H (H$^+$) number density, assuming that the
transitions are optically thin and that the electron temperature does
not strongly vary within the emitting region
\citep{gur97,ost89}. Considering a mean electron temperature in 
the shocked lobes of 10,000\,K, relative abundances of He/H\,=\,0.1
and O/H=\,3.3$\times$10$^{-4}$ \citep[appropriate for a C-rich star
like \crl,][]{lam86}, and the required atomic parameters \citep[see
e.g.][]{men83,gur97} we derive:

\begin{equation}
\label{mo1}
M_{\rm H}^{[\rm OI]}(\ms) = 1.3\times10^{-4} 
\left(\frac{n_e}{10^4 \cm3}\right)^{-1} 
\left(\frac{L_{[\rm OI]}}{0.1L_{\odot}}\right) 
\left(\frac{D}{900 {\rm pc}}\right)^2
\end{equation} \\
\begin{equation}
\label{mha}
M_{{\rm H}^+}^{\rm H\alpha}(\ms) = 9.6\times10^{-5} 
\left(\frac{n_e}{10^4 \cm3}\right)^{-1} 
\left(\frac{L_{\rm H\alpha}}{0.1L_{\odot}}\right)
\left(\frac{D}{900 {\rm pc}}\right)^2
\end{equation}

\noindent
where $L$ is the dereddened luminosity of the line and $D$ the
distance to the source. Note that Eq$.$ (\ref{mo1}) is valid for
electron densities smaller than the critical density of [\ion{O}{1}],
$n_{\rm c} \sim 10^6$\,\cm3, which is the case for \crl. On the other
hand, this equation only provides a lower limit of the total atomic
mass ($M_{\rm H}$, which includes the He contribution), since we are
assuming that most of the oxygen is neutral and in the ground state.
In deriving Eq$.$ (\ref{mha}) we have assumed the
classic radiative recombination case b for $T_{\rm
e}$\,$\sim$\,10,000\,K
\citep[i.e$.$ {\it z$_{\rm 3}$}=$n_3$/$n_{\rm e}n_{\rm H^+}$=0.25$\times$10$^{-20}$\,cm$^3$, where $n_3$ is the
population of the level $n=3$ of H;][]{gur97}.

We have separately estimated the masses of the east and west lobes
using Eqs$.$ (\ref{mo1}) and (\ref{mha}). For the east (west) lobe we
have used an average electron density of 6.5$\times$10$^{3}$\,\cm3
(9$\times$10$^{3}$\,\cm3), an average extinction of $A_V \sim 2.7$\,mag
($A_V \sim 3.1 $\,mag), and the fluxes in Table \ref{table1} 
multiplied by: ({\sc i}) a
factor 2.5, which converts the flux within a 
1\arcsec-wide slit to total
nebular flux (estimated from the direct $HST$
and ground-based images, Fig$.$\ref{f1}); and ({\sc ii}) a factor 0.5 and
0.8, for the east and west lobe respectively, that is the fraction of
the total H$\alpha$ luminosity locally produced in the lobes
\citep[not scattered emission, see \S\,\ref{features} and][]{sch81}.
We note that point ({\sc ii}) applies only to the H$\alpha$ luminosity and
not to [\ion{O}{1}] emission since the latter is not contaminated by
scattered light.  The masses derived for a distance to the source of
900\,pc are given in Table \ref{tab_mass}. In this table $M_{{\rm
H}^+}^{\rm H\alpha}$ for the east lobe is the mass of ionized shocked
gas plus the mass of ionized gas in region A' (Fig$.$\,\ref{f1}),
since the latter contributes significantly to the total H$\alpha$
luminosity. In contrast, $M_{\rm H}^{[\rm OI]}$ only accounts for the
mass of atomic shocked gas, since region A' is not detected in our
[\ion{O}{1}] long-slit spectra.


We have estimated the mass of ionized gas in region A' 
from its H$\alpha$ flux, which is $\sim$\,20\% of the total
flux (scattered + local emission) of the east lobe. The electron
density and extinction in this region is $\sim$\,10$^5$\,\cm3\ and
$A_V$$>$\,6, respectively (Sects$.$ \ref{spdist} and
\ref{sext}). Using Eq$.$ (\ref{mha}) we derive 
$M_{{\rm H}^+}^{\rm A'}$\,$>$4$\times$10$^{-5}$\,\ms. The
ionized mass in a spherical volume of gas with radius 0\farcs5,
$n_{\rm e}$\,$\sim$\,10$^5$\,\cm3, and ionization fraction $\sim$1 
(reasonable values for region A') is $\sim$\,10$^{-4}$\,\ms, also in agreement
with our previous estimate.

Subtracting the ionized mass of region A' from the total mass of the
east lobe in Table \ref{tab_mass} (column 2), the mass of ionized,
shocked gas is (3-9)$\times$10$^{-5}$\,\ms. Also, for an east
lobe-to-west lobe {\it ionized} mass ratio similar to the
corresponding {\it atomic} mass ratio, which is 1.9 (col$.$ 1 in Table
\ref{tab_mass}), $M_{{\rm H}^+}$ for the east lobe would be
4$\times$10$^{-5}$\,\ms, within the previous mass range. Note that a
similar ionization fraction for the east and west lobe is suggested by
the similar [\ion{O}{1}]/H$\alpha$ ratio in both lobes (without
including the scattered component).

We have also estimated the mass of ionized material in
\crl\ geometrically, using the mean electron densities and
the volume of the shocked lobes directly measured from the $HST$
images, assuming a cylindrical geometry for the lobe components. We
have separately considered the case of hollow lobes (with
0\farcs1-thick walls) and lobes uniformly filled with gas. Assuming
that the majority of the electrons in the nebula come from H$^+$,
i.e$.$ $n_{\rm e} \approx n_{\rm H^+}$, the ionized mass of the west
and east lobe (excluding region A') for hollow (full) cylinders is
3.5$\times$10$^{-5}$\,\ms\ (1.1$\times$$10^{-4}$\,\ms) and
4.2$\times$10$^{-5}$\,\ms\ (1.3$\times$$10^{-4}$\,\ms)
respectively. These values are consistent with those previously
obtained from the H$\alpha$ luminosity. The hollow-cylinder geometry
is a better approximation to the structure of the lobes of \crl\ than
the filled-cylinder geometry: for the latter, the derived masses
systematically exceed those obtained from the H$\alpha$ luminosity
(Table\,\ref{tab_mass}).

The high intensity of the [\ion{O}{1}] lines relative to H$\alpha$
and, in general, the observed ratios for the rest of the lines suggest
a low ionization fraction, X=$n_{\rm e}$/$n_{\rm H}$$\sim$\,0.03-0.1,
in the lobes, by comparison with the predictions of planar shock
models by \cite{har94}. These values are also consistent with the
upper limits to X derived from the ionized and atomic masses
calculated above (X$\approx$1.4$M_{{\rm H}^+}/M_{\rm H}$), X$<$0.8 and
X$<$0.6 for the west and east lobe, respectively.
An ionization fraction of only a few percent seems to be inconsistent
with the high velocity of the shocks in \crl\
($\sim$\,200\,\kms) deduced from the line width: full ionization is
expected for planar shocks with $V_{\rm s}$\,$\gsim$\,80\,\kms\
\citep{har94}. The fact that planar shocks are not 
appropriate for \crl\ could partially (but probably not totally) resolve
this problem. Fast bow-shocks produce a range of
ionization fractions along the bow-shock, with a large X value at the
head of the shock but smaller at the
wings, where the normal component of the shock velocity is
substantially smaller. The result is a smaller mean ionization
fraction than that predicted by planar shocks moving at the same
velocity, which produces equally high X values all
along the front.


\section{Discussion}
\label{discuss}

\subsection{Formation and evolution of the circumstellar envelope of CRL 618} 

It is widely accepted that the actual nebular geometry and kinematics
of \crl\ are the result of the interaction between the circumstellar
material mostly resulting from the AGB mass-loss process with fast stellar
winds ejected more recently, probably in the post-AGB phase
(\S\,\ref{intro}). Under this scenario the optical shock-excited,
rapidly expanding lobes are made of circumstellar material that is
undergoing the passage of shock fronts produced by the wind
interaction.

Our results support this scenario and add more details on how this
process is taking place. In our opinion, the morphology of the nebula,
consisting of several multiply-directed narrow lobes, suggest that the
post-AGB wind is collimated rather than spherical. The presence of
lobe components directed along different axes indicates multiple
ejection events, that could have happened with different orientations
and velocities, and/or at different epochs. The bright ripple-like
structure found along the lobes could be due to instabilities in the
flow or bow-shocks resulting from the episodic interaction between
fast collimated jets and the slowly expanding envelope ejected during
the AGB phase (see a more detailed analysis of and discussion on
$HST$ imaging in different filters in Sahai et al$.$ 2002, in preparation).

The complex spatio-kinematic structure of the lobes leads to
kinematical ages ranging from $\sim$\,100\,yr in the innermost regions
of the lobes to $\sim$\,400\,yr at the tips, without correcting for
projection effects. This age difference could be real, in which case
the innermost regions would have been more recently shocked, or it
could be due to different inclinations of, and/or different
decelerations suffered by, the lobe sub-components. Adopting a mean
inclination of 24\degr, the age of the optical nebula is
$\lsim$\,180\,yr.

The electron density does not vary systematically along the shocked
lobes of \crl\ (\S\,\ref{ne}). Since the ionization fraction is not
expected to vary drastically along the lobes (note that $V_{\rm s}$,
given by the line FWZI's, and the [\ion{O}{1}]/H$\alpha$ line ratio
have similar values all along the nebula), a roughly constant $n_{\rm
e}$ suggests that the total density is constant as well. Such a
density distribution is most likely inconsistent with most of the
[\ion{S}{2}] emission (and, probably, most forbidden lines) arising
from the lobe interior, which is presumably filled with the fast
post-AGB wind whose density is expected to decrease with the distance
(C-F$.$ Lee and Sahai, in preparation). We think that most of the
emission arises from shocked gas in thin-walled lobes, a conclusion
supported by the limb-brightened appearance of the lobes in the $HST$
images (Fig$.$\,\ref{f1}) and the better agreement between the
geometrical and H$\alpha$ masses of ionized gas derived in
\S\,\ref{mass}.

The spatial distribution of the radial optical thickness beyond the
tips of the lobes suggests a high contrast between the density of dust
in the gas filling the lobes and that in front of them, in the AGB
CSE. This is consistent with the optical lobes of
\crl\ being cavities excavated in the AGB envelope by the impact of
the post-AGB wind (but see \S\,\ref{massloss} for caveats). 

The forward shocks accelerating and shaping the lobes of \crl\ are
very likely in a radiative regime because the cooling time of the gas
immediately behind the shock front in the lobes, $t_{\rm cool}$, is
significantly smaller than the dynamical time scale of the nebula
($\lsim$\,180\,yr). Immediately after the shock, the gas temperature
is expected to rise to $\sim$\,10$^5$-10$^6$\,K for a shock speed
of $\sim$\,200\,\kms\
\cite[e.g$.$][]{dal72}. The time required by the gas to cool down 
to the present temperature of the shocked optically emitting lobes (of
the order of 10$^4$\,K) is $t_{\rm cool}$\,$\propto$\,$V_{\rm
s}^3/n_0 \Lambda(T)$, where $V_s$ is the shock velocity, $n_0$ is the
pre-shock number density of nuclei, and $\Lambda$($T$) is the cooling
rate \citep{blo90}. For a shock velocity of $V_{\rm
s}$\,$\lsim$\,200\,\kms, $\Lambda$($T$)=few$\times$10$^{-22}$
erg\,\cm3\,s$^{-1}$ is appropriate \citep{dal72} and $t_{\rm
cool}$(yr)$\approx$570$\times V_{\rm s,7}^3/n_0$, where $V_s$ and
$n_0$ are given in 10$^7$cm\,s$^{-1}$ and \cm3\ units,
respectively. The density of the molecular AGB envelope at a
radial distance of $\sim$7\arcsec, i.e$.$ ahead of the shocked
lobes, provides a lower limit to the density of the pre-shock
material since the first
interactions between the post-AGB wind and the AGB CSE 
presumably took place closer to the central star.
For a mass-loss rate
\mloss\,=\,5$\times$10$^{-5}$\,\my\ and expansion velocity 17.5\,\kms\
(\S\,\ref{intro}), $n_{\rm H_2}$\,=\,6$\times$10$^4$\,\cm3
($n_0$$\sim$2$n_{\rm H_2}$, assuming atomic pre-shock gas).  For the
numbers above, we derive $t_{\rm cool}$$<$0.04 yr. Even accounting for
an uncertainty in the preshock density of one order of magnitude, the
cooling time is much smaller than the dynamical time scale.

Thin lobe walls, where most of the shocked material seems to be
located, and high electron densities, up to $\sim$\,10$^4$\,\cm3, are
both expected in the case of radiative shocks: the low thermal pressure
after the shock passage leads to the collapse of the shocked material
in a very small and dense region after the shock \citep[see,
e.g.,][for a review]{fra99}. The large length-to-width ratio of the
lobes, $\sim$\,7, is also in better agreement with the shocks being
radiative, since for the adiabatic case the thermal pressure of the
shocked gas is very high, leading to inflated bubble-shaped lobes.

The nature of the reverse shocks, propagating in the post-AGB wind
itself towards the nebula center, is uncertain because it depends on
the unknown physical properties of the post-AGB wind. For $\rho_{\rm
AGB}/\rho_{\rm pAGB}$$\sim$\,600 (\S\,\ref{massloss}) and a reverse 
shock velocity of the order of 100\,\kms, the cooling time of the
shocked post-AGB gas would be $\lsim$\,12\,yr, and therefore the
reverse shocks would also be in a radiative regime.

In the absence of a presently active energy input, the dense gas in
the optical lobes of \crl\ should be at a temperature significantly
less than 10$^4$\,K. The high electron densities in the lobes of \crl\
would have led to a very quick recombination of the presently shocked
and ionized gas: the recombination time scale is about 10\,yr for a
total hydrogen recombination coefficient $\alpha_{\rm
H}$(10$^4$K)\,=\,4.18$\times$10$^{-13}$\,\cm3 s$^{-1}$
\citep[see, e.g.,][and references therein]{kwo00}. Also, the time for
the shocked gas to cool down one order of magnitude 
(to $\sim$\,1000\,K) is $\lsim$\,2\,yr, using a cooling rate 
$\Lambda$\,$\gsim$\,10$^{-25}$\,erg\,\cm3\,s$^{-1}$, 
appropriate for $T_{\rm e}$\,$\sim$\,10$^4$\,K and
ionization fraction $\gsim$\,0.01 \citep{dal72}.
Such a short cooling time implies that, in the absence of a continuous
ionization/heating source, the gas in the lobes of \crl\ should have
become atomic and cool in a time scale much smaller than the age of
nebula, $\lsim$\,180\,yr. Even if we have overestimated its
kinematical age,
\crl\ is at least about 30 years old, since it was observed
with similar characteristics in 1973 by, e.g., \cite{wes75}. The most
likely energy source for maintaining a high temperature ($>$1000\,K)
in the lobes is an ongoing interaction between the post-AGB wind and
the circumstellar material (note that the central star is not hot
enough to ionize and and heat the lobes). This implies that the fast
wind, which is ultimately responsible for the shaping and evolution of
its circumstellar envelope, is currently active in \crl.

This fast wind is not necessarily stationary, but
may rather be pulsed (episodic) with a period smaller but comparable
to the recombination time scale of the shocked gas
($\sim$\,10\,yr). In this case, the shocked material (initially fully
ionized) would have time to partially recombine between pulses, which
might be a contributing factor to maintaining a low ionization
fraction in the lobes of \crl\ in spite of the high velocity of the
shocks. On the other hand, to keep the temperature of the shocked gas
above $T_{\rm e}$\,$>$\,$10^3$\,K, the pulse period should be
$>$\,2\,yr. The idea of a pulsating fast wind is consistent with the
presence of the multiple bright, bow-shaped and ripple-like structures
seen along the lobes of \crl\ (Fig$.$\,\ref{f1}) since their
separations correspond to time-scales of $\sim$\,10\,yr.

Finally, we have reported a recent increase (decre\-ase) of the
intensity of the [\ion{O}{3}]$\lambda$5007\,\AA\
([\ion{O}{1}]$\lambda$6300\,\AA) lines in \crl\
(\S\,\ref{fluxes}). This change also provides indirect evidence for 
the presence of continuously evolving shocks resulting from a continuous
interaction between the ongoing post-AGB wind and the surrounding
media. Since shocks are the main excitation mechanism of the gas
in the lobes of CRL\,618, where most of the forbidden line emission is
detected, the observed [\ion{O}{3}] brightening probably has its
origin in a change of the shock properties (rather than in the stellar
radiation). The [\ion{O}{3}]/H$\beta$ ratio is very sensitive to the
shock velocity (\hrh) and therefore, the recent increase of the
[\ion{O}{3}] flux may indicate that the ongoing shocks are faster now
than 10 years ago. As a consequence of the shock acceleration
the ionization fraction of the lobes could have increased, leading to
a weakening of the [\ion{O}{1}] lines. The increasing velocity of the
shocks could point to acceleration of the stellar post-AGB wind. The
time scale of the [\ion{O}{3}]/[\ion{O}{1}] brightening/weakening
(observed over $\gsim$\,10\,yr) is consistent with the expected shock
velocity and the size of the shocked-gas emitting region: a
90-200\,\kms\ shock will cross the 0\farcs1-wide lobe walls
($\sim$1.3$\times$10$^{15}$\,cm) in only 4.7-2.1\,yr.

\subsection{Future circumstellar evolution of \crl}

The shocked optical lobes of \crl\ could evolve into 
a fast, bipolar molecular outflow if the time scale of molecular
reformation in the shocked gas is smaller than the time required by
the expanding ionization front (produced by the central star) to reach
and fully ionize the whole nebula.

The present radiative, forward shocks will continue propagating
through the molecular cocoon (traced by CO), accelerating and shaping
more and more material of the AGB CSE. We are observing the beginning
of this process, when the shocked material is still hot (and,
therefore, primarily visible in the optical), and its mass is still
low compared to that of the unshocked envelope.
The hot shocked material will cool down, allowing molecule
reformation, and would be observed as an extended, bipolar outflow
moving at high velocity.
This is what presumably already happened to a part of the shocked
material: the fast molecular bipolar outflow observed in the innermost
parts of CRL\,618 (\S\,\ref{intro}, discussed in detail in S\'anchez
Contreras et al$.$ 2002 --- paper II, in preparation). If the shocks
have sufficient energy to accelerate most of the AGB envelope, the mass
of the future fast, bipolar outflow of CRL\,618 could be a significant
fraction of the total mass of the AGB CSE, $\sim$\,1.5\,\ms. The
interaction of this fast and massive molecular jet with the
interstellar medium or the low density, outermost parts of the
CRL\,618 CSE could also lead to new shocks that
would be observed, in their first stages, through optical line
emission surrounding the molecular bipolar outflow.

In our opinion, the possible future evolution of \crl\ described above
is what already occurred in another well studied protoplanetary
nebula, OH\,231.8+4.2. This object shows a very collimated and fast
molecular outflow which was presumably shaped and accelerated by
radiative shocks resulting from a two-wind interaction process
$\sim$\,1000\,yr ago \citep{san00,alc01}; this outflow should have
been in the past a very intense emitter of recombination and forbidden
lines at short wavelengths, as the optical lobes in \crl. Surrounding
the collimated molecular outflow of OH\,231.8+4.2, there is a wide
bow-shaped nebula of shock-excited warm gas which is observed in the
visible \citep{buj02,san00}. This component is believed to result from
the hydrodynamical interaction between the bipolar CO outflow and the
tenuous circumstellar material, the shocks generated in this
interaction being more likely in an adiabatic regime.

The central star of OH\,231.8+4.2 is still very cool ($\sim$\,2500\,K)
and therefore is not able to ionize the circumstellar material. The
evolution of the optical lobes of \crl\ into a fast molecular
outflow (like in OH\,231.8+4.2) in the future requires 
that once the wind interaction stops, the time required by the
ionization front, produced by the hot central star, to reach the lobes ($t_{\rm
ion}$) be larger than the molecular reformation time scale ($t_{\rm
mol}$).

In the relatively dense envelopes of PPNe, the time scale for molecule
formation is $\lsim$\,1000\,yr, as deduced from the widespread
existence of fast ($V\gsim$\,100\,\kms) molecular outflows with
kinematical ages $\lsim$\,10$^3$\,yr
\cite[see, e.g.,][]{buj01}: note that molecules are totally destroyed by 
shocks at speeds $\gsim$30\,\kms\ \citep{hol89} and therefore
molecules moving at such high velocities have most likely been
reformed in the post-shock gas. The kinematical age of the compact
molecular outflow of \crl\ itself, moving at $\gsim$100\,\kms, is only
of the order of 10$^2$\,yr (\S\,\ref{intro}), so, in general, we can
expect molecule reformation to occur in $t_{\rm
mol}$\,$\sim$\,10$^2$-10$^3$\,yr. This value is also in reasonable
agreement with theoretical estimates of $t_{\rm mol}$ in dense,
$n_{\rm H}$\,$\sim$\,10$^6$\,\cm3, post-shock gas
\citep{hol89}.


The time required by the ionization front to reach the lobes, $t_{\rm
ion}$, depends on the relative expansion velocity of the front and the
lobes. The expansion velocity of the ionization front in \crl,
estimated from radio continuum mapping with high-angular resolution in
different epochs, has decreased with time: from
50-74($D$/0.9\,kpc)\,\kms\ \citep{kwo81} to 26($D$/0.9\,kpc)\,\kms\
\citep{spe83}. Expanding at this rate, 
the ionization front will never reach the lobes, which are moving much
faster (up to 200\,\kms), and the molecules will reform in the
shocked gas sooner or later. If the shocked lobes are decelerated by
interaction with the surrounding circumstellar gas and/or the velocity
of the ionization front increases (due, e.g., to the evolution of the
central star, which will become hotter and hotter, and the dilution of
the circumstellar gas) then the UV stellar photons will be able to
fully ionize the lobes of \crl. The ionization time-scale depends on
different unknown physical parameters of the central star, such as
mass, temperature, mass-loss rate, etc, and their evolution with
time. The shocked gas would be observed as a molecular outflow during
$t_{\rm ion}-t_{\rm mol}$, whenever $t_{\rm ion}> t_{\rm mol}$,
otherwise \crl\ will never develop a massive bipolar molecular
outflow (like e.g$.$ OH\,231.8+4.2).


\section{Conclusions}
\label{conc}

We have obtained ground-based, long-slit spectra in the ranges
[6233-6806]\,\AA\ and [4558-5331]\,\AA\ for three different (1\arcsec-wide)
slit positions in the PPN \crl. Based 
on the analysis of these spectra and direct (H$\alpha$ and continuum)
images, we find the following results: 

$-$ The Balmer line emission results from the superposition of
unscattered and scattered components. The unscattered emission is
locally produced in the shocked lobes and in a inner, compact region
labeled A' (Fig$.$\,\ref{f1}). The scattered radiation originates in
the central \ion{H}{2} region (and also maybe in region A') and is
reflected by the dust in the shocked lobes and in the unshocked AGB
envelope. Most forbidden lines are entirely produced in the
lobes but there is also forbidden line emission (e.g$.$ [\ion{Fe}{3}]
and [\ion{S}{3}]) arising in the \ion{H}{2} region and region A'.


$-$ We identify the Wolf-Rayet bump at $\sim$\,6540\,\AA\ superimposed
on the scattered continuum. This bump, and a number of high-excitation
transitions such as, e.g.,
\ion{He}{2}$\lambda$4686\,\AA, originate in the dense
($n_{\rm e}$\,$\sim$\,10$^5$-10$^7$\,\cm3) inner \ion{H}{2} region,
and are seen after being reflected by the nebular dust.

$-$ The spectrum of region A' is similar to that of the inner
\ion{H}{2} region (dominated by high-excitation lines) but remarkably
different from the spectrum of the shocked lobes (dominated by
low-excitation lines), and is consistent with $n_{\rm
e}$\,$\sim$10$^4$-10$^5$\,\cm3 and full ionization in this
region. Thus, region A' very likely represents the outermost parts of
the central
\ion{H}{2} region, i.e$.$ it is gas ionized by the UV stellar
radiation rather than by shocks.

$-$ The shock velocity derived from the profiles of line emission
arising in the shocked lobes is $\gsim$\,200\,\kms.  However, the line
ratios in the lobes are consistent with the predictions of bow-shock
models with a lower shock velocity (less than about 90\,\kms, and 
probably $\sim$\,75\,\kms).

$-$ We have measured an increase (decrease) of the
[\ion{O}{3}]$\lambda$5007\AA\ ([\ion{O}{1}]$\lambda$6300\AA) line flux
by a factor $\sim$\,2.5 (0.7) with respect to observations performed
$\sim$\,10\,yr ago. Such variations could be due to
a recent increase of the shock velocity. 

$-$ The mean nebular inclination (with respect to the plane of the
sky) is $i$=24\degr$\pm$6\degr\ and, very likely, less than 39\degr.

$-$ The material in the lobes of \crl\ is moving away from the star at
velocities up to $\sim$\,200\,\kms. The projected velocity of the gas
shows a non-linear radial variation, resulting in a
kinematical age that varies from $\sim$100\,yr near the nebular
center to $\sim$\,400\,yr at the tips, without correcting projection
effects. For a mean inclination of $i$=24\degr, the kinematical age of
the optical lobes is $\lsim$\,180\,yr.



$-$ The inner \ion{H}{2} region and region A' seem to be 
expanding at moderate velocity, $\lsim$\,30\,\kms, 
much slower than the shocked lobes.

$-$ The extinction along the nebula is found to
decrease from the center (A$_V$\,$>$\,10\,mag) to the lobe tips
($\sim$\,2\,mag); beyond the bright optical lobes, where an H$\alpha$
scattered halo is visible, the extinction increases again. This increase
indicates a large density contrast between the dust inside the
optical lobes and beyond them, i.e$.$ in the AGB CSE unaltered by
the shocks. This result is consistent with the
lobes of \crl\ being cavities in the AGB CSE excavated by
post-AGB winds.

$-$ The electron density, with mean values $n_{\rm
e}$\,$\sim$\,[5-8]$\times$10$^3$ and
$\sim$\,8$\times$10$^3$-10$^4$\,\cm3, for the east and west lobe,
respectively, shows no systematic variation along the lobes.

$-$ The ionization fraction in the lobes is X$\sim$\,0.03-0.1, derived
by comparison of different line ratios with predictions from
planar shocks models. The total number density in the lobes is
then $\sim$10$^5$-10$^6$\,\cm3.

$-$ The mass of atomic gas in the east (west) lobe is
$>$1.3$\times$10$^{-4}$\,\ms\ ($>$7.0$\times$10$^{-5}$\,\ms).  The
masses of ionized gas in the lobes are smaller:
$\sim$6$\times$10$^{-5}$\,\ms, for the east lobe, and
$\sim$4$\times$10$^{-5}$\,\ms\ for the western lobe. The ionized mass
in region A' is comparable to the ionized mass in the lobes or even
higher, up to $\sim$10$^{-4}$\,\ms.

$-$ The limb brightening of the optical lobes and the lack of a
systematic spatial gradient in $n_{\rm e}$ suggest that most of the
line emission arises in a thin shell of shocked material rather than
in gas filling the interior of the lobes. The thin-shell geometry is
expected from radiative shocks shaping and accelerating the AGB
circumstellar envelope. The radiative nature of the shocks is
inferred from the short cooling time of the immediate post-shock
gas, $\lsim$\,0.04\,yr (relative to the dynamical time).

$-$ The post-AGB wind is probably currently active, shocking
and heating the AGB circumstellar material, since the time required
for the gas in the shocked lobes (presently partially ionized) to cool
down significantly below 10$^4$\,K is $\lsim$\,2\,yr, much shorter
than the age of the nebula.




\acknowledgments 
The authors are grateful to Noam Soker for reading and commenting on
this paper, to A$.$ Castro-Carrizo for helping during the long-slit
observations, and to C-F$.$ Lee for fruitful conversations during the
writing of this paper. This work was performed at the Jet Propulsion
Laboratory, California Institute of Technology, under a contract with
the National Aeronautics and Space Administration and has been
partially supported by a NASA Long Term Space Astrophysics grant to
RS. This research has made use of the USNOFS Image and Catalog Archive
operated by the United States Naval Observatory, Flagstaff Station
(http://www.nofs.navy.mil/data/fchpix/). The authors also acknowledge
the use of NASA's Astrophysical Data System Abstract Service (ADS).

\clearpage

\begin{deluxetable}{lccrc} 
\tabletypesize{\small}
\tablecaption{Observed line fluxes for CRL 618} 
\tablewidth{0pt} 
\tablehead{ 
\colhead{Line}    &  \colhead{Wavelength} & \multicolumn{3}{c}{Flux} \\ 
\colhead{} & \colhead{(\AA)}  & \multicolumn{3}{c}{(10$^{-15}$ erg\,s$^{-1}$\,cm$^{-2}$)} \\
\cline{3-5} 
\colhead{} & \colhead{} & \colhead{West} & \colhead{} & \colhead{East}}
\startdata 
$[$\ion{Mg}{1}$]$ & 4571.1 & 1.2(0.3) & & 3.5(0.9) \\ 
W-R bump & $\sim$\,4620-55\tablenotemark{a} & -- & & 6(2) \\
$[$\ion{Fe}{3}$]$, \ion{C}{4} & 4658.05,4658.30\tablenotemark{b} & -- & & 3.7(0.8) \\
\ion{He}{2}        & 4686.55 & -- & & 0.8(0.2) \\
$[$\ion{Fe}{3}$]$  & 4701.5  & -- & & 1.7(0.4) \\
$[$\ion{Fe}{3}$]$  & 4733.9  & -- & & 0.5(0.4) \\
$[$\ion{Fe}{3}$]$  & 4754.7  & -- & & 0.5(0.4) \\
H$\beta$ & 4861.3 & 11(1)  & & 46(4) \\
$[$\ion{O}{3}$]$ & 4958.9  & 0.4(0.2) & & 2(1) \\
$[$\ion{O}{3}$]$ & 5006.8  & 1.3(0.2) & & 6(1) \\
$[$\ion{Fe}{3}$]$ & 5011.3 & -- & & 1(0.6) \\
\ion{He}{1} & 5015.7       & --  &     & 0.6(0.7) \\
$[$\ion{Fe}{2}$]$ & 5158.0,5158.8\tablenotemark{b} & 1.5(0.2) & & 2.5(0.7) \\
$[$\ion{N}{1}$]$ & 5197.9 & 9.5(0.3) & & 23(1) \\
$[$\ion{N}{1}$]$ & 5200.3 & 5.8(0.3) & & 11.0(0.7) \\
$[$\ion{Fe}{2}$]$ & 5261.6 &  0.9(0.3) & & 1.2(0.6) \\
$[$\ion{Fe}{3}$]$ & 5270.4 & -- & & 4.3(0.6) \\
$[$\ion{O}{1}$]$ & 6300.3 & 55(3) & & 94(4) \\
$[$\ion{S}{3}$]$ & 6312.1 & -- & & 2.5(0.6) \\
$[$\ion{O}{1}$]$ & 6363.8 & 18(2) & & 33(3) \\
$[$\ion{N}{2}$]$ & 6548.0 & 10(2) & & 16(2) \\
H$\alpha$ & 6562.8 & 83(4) & & 390(8) \\
$[$\ion{N}{2}$]$  & 6583.5 & 33(3) & & 56(3) \\
\ion{He}{1} & 6678.2 & -- & & 1.8(0.8) \\
$[$\ion{S}{2}$]$ & 6716.4 & 18(2) & & 34(3) \\
$[$\ion{S}{2}$]$ & 6730.8 & 35(2) & & 60(3) \\
\enddata 

\tablenotetext{a}{\ion{N}{5}\,$\lambda$4620, \ion{N}{3} 
$\lambda$$\lambda$4634,4641,\ion{C}{3} $\lambda$$\lambda$4647,4650}
\tablenotetext{b}{Blend of lines}
\tablecomments{Errors are given in parenthesis.}
\label{table1}
\end{deluxetable}




\begin{table}
\small
\begin{center}
\caption{[\ion{O}{1}]$\lambda$6300/H$\alpha$ line ratios for \crl}
\label{oivstime}
\begin{tabular}{lcccc}
\tableline\tableline
\multicolumn{1}{c}{Observation} & J.D. & \multicolumn{2}{c}{$[$\ion{O}{1}$]$/H$\alpha$} & Refs. \\
\multicolumn{1}{c}{Date}       &   (days)  &   East &  West &  \\
\tableline
1973 Oct-1974 Feb  &   2442032 &   1.2\tablenotemark{a}  & 1.2\tablenotemark{a} & 1  \\ 
1979 Dec    &   2444223 &   0.9  &   1.3  & 2 \\ 
1989 Feb-1990 Oct & 2447846 &   0.4  &   0.8  & 3 \\
1990 Oct-1990 Nov   &    2448197 &   0.5  &   0.3  & 4 \\
1994 Oct    &    2449641 &   0.4  &   1.1  & 5 \\
2000 Nov    &    2451864 &   0.2  &   0.7  & 6\\
\tableline
\end{tabular}
\tablenotetext{a}{Ratio obtained for the whole nebula}
\tablerefs{(1) Westbrook et al$.$ 1975; (2) Schmidt \& Cohen 1981;
(3) Kelly et al$.$ 1992; (4) Trammell et al$.$ 1993; 
(5) Baessgen et al$.$ 1997; (6) This work.}
\end{center}
\end{table}

\begin{table*}
\begin{center}
\caption{Velocity data from the [\ion{O}{1}]$\lambda$6300\,\AA\ line}
\small
\label{table2}
\begin{tabular}{rccccccc}
\tableline\tableline
& \multicolumn{3}{c}{slit N93} & & \multicolumn{3}{c}{slit S93} \\
\cline{2-4}
\cline{6-8}
offset & $V_{\rm LSR}$+21 & FWHM\tablenotemark{a} & FWZI\tablenotemark{b} &
& $V_{\rm LSR}$+21 & FWHM\tablenotemark{a}  & FWZI\tablenotemark{b} \\
\multicolumn{1}{c}{($''$)} & (\kms)  & (\kms) & (\kms) & & (\kms) & (\kms)  & (\kms) \\
\cline{2-4}
\cline{6-8}
\multicolumn{8}{l}{West (weak) lobe} \\
$-$6\farcs2   &  76 & 125 & 220 && 56 & 136 & 190 \\
$-$3\farcs7   &  56 &  75 & 190 && 44 & 73 & 190 \\
$-$1\farcs0   &  61 &  61 & 90  && 56 & 61 & 100 \\
\multicolumn{8}{l}{East (bright) lobe}\\
1\farcs0 & $-$49 & 150   & 200 &&  -60 & 125 & 170\\
3\farcs2 & $-$31 & 110   & 250 &&  -45 & 125 & 240 \\
5\farcs4 & $-$39 & 117   & 230 &&  -54 & 146 & 230 \\
6\farcs5 & $-$69 & 138   & 210 &&  -64 & 177 & 220 \\ 
\tableline
\end{tabular}
\tablenotetext{a}{Deconvolved with the spectral resolution, $\sim$\,50\,\kms.}
\tablenotetext{b}{Measured at a $\sim$\,3$\sigma$ level.}
\end{center}
\end{table*}

\begin{table}
\begin{center}
\caption{Atomic and ionized mass in \crl}
\label{tab_mass}
\begin{tabular}{lll}
\tableline\tableline
        & $M_{\rm H}^{[\rm OI]}$ & $M_{{\rm H}^+}^{\rm H\alpha}$\\ 
        & (\ms)                  &  (\ms)                       \\
\tableline
East & $>$1.3$\times$10$^{-4}$ & 1.3$\times$10$^{-4}$\tablenotemark{(a)} \\ 
West & $>$7.0$\times$10$^{-5}$ & 4.0$\times$10$^{-5}$ \\
\tableline
\end{tabular}
\tablenotetext{(a)}{Mass of region A' + east lobe
}
\end{center}
\end{table}

\clearpage


\begin{figure}    
\caption{Direct imaging of the PPN \crl. {\it a)} 
Narrow-band, ground-based image of the continuum emission at
6616\,\AA. Contours are 3$\sigma$, 5$\sigma$, 10$\sigma$, 15$\sigma$,
and from 20$\sigma$ to 120$\sigma$ by 20$\sigma$, with
$\sigma$=7$\times$10$^{-19}$\,\bri; {\it b)} Narrow-band, ground-based
image of the H$\alpha$ emission line (continuum subtracted). Contours
are 5$\sigma$, 10$\sigma$, 20$\sigma$, 40$\sigma$, 150$\sigma$, and
from 300$\sigma$ to 2300$\sigma$ by 500$\sigma$, with
$\sigma$=6.5$\times$10$^{-17}$\,\brib; {\it c)} Narrow-band
WFPC2/$HST$ image in the light of H$\alpha$+continuum.  The continuum
contribution in this image is less than $\sim$\,3\% in any region. (PLEASE, DOWNLOAD csanchez.fig1.gif).}
\label{f1}
\end{figure}

\begin{figure}    
\caption{ 
Long-slit spectra of selected emission lines in \crl\ observed with
grating R1200Y for slits N93 (top), S93 (middle), and PA3
(bottom). The $HST$ and ground-based H$\alpha$ image of the nebula are
shown in the left-most panels (top and bottom, respectively). Slits
used for spectroscopy are superimposed on the ground-based
image. Contours in the spectra are: 3$\sigma$$\times$1.8$^{(i-1)}$
from $i$=1 to 15 with $\sigma$=10$^{-16}$\bri\ for N93 and
$\sigma$=1.4$\times$10$^{-16}$\bri\ for S93 and PA3. The LSR velocity
scale (in \kms) is shown over the N93 H$\alpha$ spectra. The systemic
velocity ($\sim-$21\,\kms) is indicated by the vertical line. (PLEASE, DOWNLOAD csanchez.fig2.gif).} 
\label{f2}
\end{figure}

\begin{figure}
\caption{Smoothed long-slit spectra of [\ion{S}{3}]$\lambda$6312.1\,\AA\ and
\ion{He}{1}$\lambda$6678.2\,\AA\ for slit N93. The total velocity range 
is 600\,\kms\ as in Fig$.$\,\ref{f2}. Contours are:
2$\sigma$$\times$1.5$^{(i-1)}$ from $i$=1 to 15 and with
$\sigma$=4$\times$10$^{-17}$\bri. The vertical line is placed at LSR
\vsys\ ($\sim-$21\,\kms). Feature ``a'' is indicated (see also
Fig$.$\ref{f4}). (PLEASE, DOWNLOAD csanchez.fig3.gif).}
\label{f3}
\end{figure}

\begin{figure}    
\caption{Long-slit spectra of H$\alpha$, [\ion{N}{2}]$\lambda$6584\AA\ 
(color/grey scale and black contours), and [\ion{O}{1}]6300 emission
(green/light-grey contours) for slit N93 (top), S93 (middle), and PA3
(bottom). The intensity levels represented are the same for all
transitions. Note that except for the presence of features A, B, and C
in the H$\alpha$ spectrum, the profiles and fluxes of H$\alpha$ and
[\ion{O}{1}] are comparable. In the [\ion{N}{2}] profile, feature
``a'' is the counterpart to feature A. (PLEASE, DOWNLOAD csanchez.fig4.gif).}
\label{f4}
\end{figure}

\begin{figure}[ht]    
\caption{Long-slit spectra in the blue (grating R900V) 
obtained along the axis of CRL 618 (position N93 in
Fig$.$\ref{f2}). In the top panel, we plot the total spectrum with
degraded spatial and spectral resolutions by a factor 70 and 50\%,
respectively, with respect to the nominal value in order to show the
continuum and weakest emission lines. In the lower panel, long-slit
spectra for selected lines are shown.  As in Fig$.$\,\ref{f2}, these
spectra have been smoothed within a 3$\times$3 pixel box with no
significant degradation (less than 4 percent) of the nomimal spatial
and spectral resolution (see \S\,\ref{obsop}). (PLEASE, DOWNLOAD csanchez.fig5.gif).}
\label{f5}
\end{figure}

\begin{figure}[ht]     
\caption{{\it Top)} Spatial variation 
along slit N93 of the optical depth at 4861\,\AA\ (black squares),
derived from the H$\alpha$-to-H$\beta$ relative flux (the H$\alpha$
and H$\beta$ spatial profiles are drawn using dashed
lines). Light (blue) solid lines represent the radial optical depth,
$\tau_r$, due to dust in the AGB CSE and the post-AGB wind (see
\S\,\ref{sext}). The thick (red) line is the sum of these two optical
depths. {\it Middle)} [\ion{S}{2}]$\lambda$6716/$\lambda$6731 doublet
ratio, which is an indicator of the electron density, along N93 (black
squares). The spatial profiles of the
[\ion{S}{2}]$\lambda$$\lambda$6716,6731 doublet (dashed lines) are
also shown. {\it Bottom)} $HST$ (colour) and ground-based (contours)
images of \crl. (PLEASE, DOWNLOAD csanchez.fig10.gif).}
\label{f10}
\end{figure}


\begin{thebibliography}{}

\bibitem[Alcolea et al.(2001)]{alc01} Alcolea, J., 
Bujarrabal, V., S{\'a}nchez Contreras, C., Neri, R., \& Zweigle, J.\ 2001, 
\aap, 373, 932 

\bibitem[Aller \& Keyes(1987)]{all87} Aller, L.~H.~\& Keyes, 
C.~D.\ 1987, \apjs, 65, 405. 


\bibitem[Bachiller et al.(1988)]{bac88} Bachiller, R., G\'omez-Gonz\'alez, J., Bujarrabal, V., \& Mart\ai n-Pintado, J.\ 1988, \aap, 196, L5


\bibitem[Blondin, Fryxell, \& Konigl(1990)]{blo90} Blondin, J.~M., 
Fryxell, B.~A., \& Konigl, A.\ 1990, \apj, 360, 370

\bibitem[Bujarrabal et al.(1988)]{buj88} Bujarrabal, V., Gomez-Gonzalez, J., Bachiller, R., \& Martin-Pintado, J.\ 1988, \aap, 204, 242

\bibitem[Bujarrabal et al.(2001)]{buj01} Bujarrabal, V., Castro-Carrizo, A., Alcolea, J., \& S{\' a}nchez Contreras, C.\ 2001, \aap, 377, 868


\bibitem[Bujarrabal et al.(2002)]{buj02} Bujarrabal, V., Alcolea, J., S{\' a}nchez Contreras, C., and Sahai, R., 2002, A\&A in press

\bibitem[Cant{\' o} et al.(1980)]{can80} 
Cant{\' o}, J., Meaburn, J., Theokas, A.~C., \& Elliott, K.~H.\ 1980,
\mnras, 193, 911

\bibitem[Cardelli, Clayton, \& Mathis(1989)]{car89} 
Cardelli, J.~A., Clayton, G.~C., \& Mathis, J.~S.\ 1989, \apj, 345, 245







\bibitem[Calvet and Cohen(1978)]{cal78} Calvet, N.\ and Cohen, M.\ 1978, 
\mnras\ 182, 687

\bibitem[Carsenty and Solf(1982)]{car82} Carsenty, U.\ and Solf, J.\ 1982, 
\aap, 106, 307

\bibitem[Cernicharo et al$.$(1989)]{cer89} Cernicharo, J., Gu\'elin, M., Martin-Pintado, J., Pe\n alver, J., and Mauersberger, R., 1989 A\&A 222, L1


\bibitem[Dalgarno \& McCray(1972)]{dal72} Dalgarno, A.~\& McCray, R.~A.\ 1972, \araa, 10, 375



\bibitem[Frank(1999)]{fra99} Frank, A.\ 1999, New Astronomy Review, 43, 31 




\bibitem[Gurzadyan(1997)]{gur97} Gurzadyan, G.~A.\ 1997, 
The Physics and Dynamics of Planetary Nebulae, Springer-Verlag Berlin
Heidelberg New York.~ Also Astronomy and Astrophysics Library.

\bibitem[Gammie et al.(1989)]{gam89} Gammie, C.~F., Knapp, G.~R., Young, K., Phillips, T.~G., \& Falgarone, E.\ 1989, \apjl, 345, L87

\bibitem[Goodrich(1991)]{goo91} Goodrich, R.W.\ 1991, \apj\ 376, 654

\bibitem[Gottlieb and Liller(1976)]{got76} Gottlieb, E.W.\ and Liller, Wm.\
1976, \apj\ 207, L135


\bibitem[Hajian, Phillips, \& Terzian(1996)]{haj96} Hajian, 
A.~R., Phillips, J.~A., \& Terzian, Y.\ 1996, \apj, 467, 341. 

\bibitem[Hartigan, Raymond, and Hartmann(1987)]{har87} Hartigan, P., 
Raymon, J., and Hartmann, L.\ 1987, \apj\ 316, 323


\bibitem[Hartigan, Morse, \& Raymond(1994)]{har94} Hartigan, P., Morse, J.~A., \& Raymond, J.\ 1994, \apj, 436, 125


\bibitem[Hollenbach \& McKee(1989)]{hol89} Hollenbach, D.~\& McKee, 
C.~F.\ 1989, \apj, 342, 306

\bibitem[Howarth(1983)]{how83} Howarth, J.D.\ 1983, \mnras\ 203, 301


\bibitem[Kastner, Soker, and Rappaport(2000)]{kas00} Kastner, J., Soker, N., and Rappaport, S.A., 2000, proceedings of the ASP 
Conf.~Ser.~199: Asymmetrical Planetary Nebulae II: From Origins to
Microstructures


\bibitem[Kaler(1978)]{kal78} Kaler, J.~B.\ 1978, \apj, 220, 887


\bibitem[Kelly, Latter, and Rieke(1992)]{kel92} Kelly, D.M., Latter, W.B., 
and Rieke, G.H.\ 1992, \apj\ 395, 174

\bibitem[Knapp \& Morris(1985)]{kna85} Knapp, G.~R.~\& Morris, M.\ 1985, \apj, 292, 640

\bibitem[Knapp, Sandell, \& Robson(1993)]{kna93} Knapp, 
G.~R., Sandell, G., \& Robson, E.~I.\ 1993, \apjs, 88, 173. 

\bibitem[Kwok \& Feldman(1981)]{kwo81} Kwok, S.~\& Feldman, P.~A.\ 1981, \apjl, 247, L67

\bibitem[Kwok \& Bignell(1984)]{kwo84} Kwok, S.~\& Bignell, R.~C.\ 1984, \apj, 276, 544

\bibitem[Kwok(2000)]{kwo00} Kwok, S.\ 2000, The origin and evolution of planetary nebulae / Sun Kwok.~Cambridge ; New York : Cambridge University Press, 2000.~(Cambridge astrophysics series ; 33), 

\bibitem[Lambert et al.(1986)]{lam86} Lambert, D.~L., Gustafsson, B., Eriksson, K., \&
Hinkle, K.~H.\ 1986, \apjs, 62, 373




\bibitem[Leuenhagen, Hamann, \& Jeffery(1996)]{leu96} 
Leuenhagen, U., Hamann, W.-R., \& Jeffery, C.~S.\ 1996, \aap, 312, 167 







\bibitem[Meixner et al.(1998)]{mei98} 
Meixner, M., Campbell, M.~T., Welch, W.~J., \& Likkel, L.\ 1998, \apj, 509, 
392. 

\bibitem[Mendoza(1983)]{men83} Mendoza, C.\ 1983, IAU Symp.~103: 
Planetary Nebulae, 103, 143


\bibitem[Mart\ai n-Pintado et al.(1988)]{mar88} Mart\ai n-Pintado, 
J., Bujarrabal, V., Bachiller, R., G\'omez-Gonz\'alez, J., \& Planesas, P.\ 
1988, \aap, 197, L15 


\bibitem[Mart\ai n-Pintado et al.(1993)]{mar93} Mart\ai n-Pintado, J., Gaume, R., Bachiller, R., \& Johnson, K.\ 1993, \apj, 419, 725

\bibitem[Neri et al$.$(1992)]{ner92} Neri, R., Garc{\'{\i}}a-Burillo, S., 
Gu\`elin, M., Guilloteau, S., and Lucas, R.\ 1992, A\&A 262, 544





\bibitem[Osterbrock(1989)]{ost89} Osterbrock D.E., 1989, 
Astrophysics of Gaseous Nebulae and Active Galactic Nuclei. 
University Science Books, Mill Valley, CA 

\bibitem[Phillips et al.(1992)]{phi92} Phillips, J.~P., Williams, P.~G., Mampaso, A., \& Ukita, N.\ 1992, \aap, 260, 283



\bibitem[Riera, Phillips, \& Mampaso(1990)]{rie90} Riera, A., Phillips, J.~P., \& Mampaso, A.\ 1990, \apss, 171, 231 

\bibitem[Reipurth et al.(2000)]{rei00} Reipurth, B., Yu, K., 
Heathcote, S., Bally, J., \& Rodr{\'i}guez, L.~F.\ 2000, \aj, 120,
1449.

\bibitem[Schmidt and Cohen(1981)]{sch81} Schmidt, G.D.\ and Cohen, M.\ 
1981, \apj\ 246, 444

\bibitem[Sahai \& Trauger(1998)]{sah98} Sahai, R.~\& Trauger, J.~T.\ 1998, \aj, 116, 1357 





\bibitem[S{\'a}nchez Contreras et al(2000)]{san00} S{\'a}nchez 
Contreras, C., Bujarrabal, V., Miranda, L.~F., \& Fern{\'a}ndez-Figueroa, 
M.~J.\ 2000, \aap, 355, 1103 



\bibitem[Spergel, Giuliani, \& Knapp(1983)]{spe83} Spergel, 
D.~N., Giuliani, J.~L., \& Knapp, G.~R.\ 1983, \apj, 275, 330

\bibitem[Schwarz et al.(1997)]{sch97} Schwarz, H.~E., Aspin, C., Corradi, R.~L.~M., \& Reipurth, B.\ 1997, \aap, 319, 267 

\bibitem[Trammell, Dinerstein, and Goodrich(1993)]{tra93} Trammell, S.R., 
Dinerstein, H.L., and Goodrich, R.W.\ 1993, \apj\ 402, 249

\bibitem[Trammell(2000)]{tra00} Trammell, S.~R.\ 2000, ASP 
Conf.~Ser.~199: Asymmetrical Planetary Nebulae II: From Origins to 
Microstructures, 147 

\bibitem[Tylenda, Acker, \& Stenholm(1993)]{tyl93} Tylenda, R., Acker, A., 
\& Stenholm, B.\ 1993, \aaps, 102, 595


\bibitem[Welch et al.(1999)]{wel99} Welch, C.~A., Frank, A., 
Pipher, J.~L., Forrest, W.~J., \& Woodward, C.~E.\ 1999, \apjl, 522, L69. 

\bibitem[Westbrook et al$.$(1975)]{wes75} Westbrook, W.E., Becklin, E.E., Merrill, K.M., Neugebauer, G., Schmidt, M., Willner, S.P., and Wynn-Williams, C.G.\ 1975, \apj\ 202, 407


\end{thebibliography}
\end{document}